\documentstyle[aps]{revtex}
\input epsf.tex

\def\spose#1{\hbox to 0pt{#1\hss}}
\def\lta{\mathrel{\spose{\lower 3pt\hbox{$\mathchar"218$}}
     \raise 2.0pt\hbox{$\mathchar"13C$}}}
\def\gta{\mathrel{\spose{\lower 3pt\hbox{$\mathchar"218$}}
     \raise 2.0pt\hbox{$\mathchar"13E$}}}

\def\eg{{\it e.g.,\,}}
\def\ie{{\it i.e.,\,}}
\def\etal{{\it et al.\,}}

\def\half{{\textstyle{1\over2}}}
\def\be{\begin{equation}}
\def\ee{\end{equation}}
\def\bea{\begin{eqnarray}}
\def\eea{\end{eqnarray}}
\def\lbl{\label}
\def\hm{{\cal H}^3}

\def\xic{\xi^{c}_\Phi({\bf x}, {\bf x}^\prime)}
\def\xico{\xi^{c}_\Phi({\bf x}, {\bf x})}
\def\xiu{\xi^{u}_\Phi({\bf x}, {\bf x}^\prime)}
\def\xiug{\xi^{u}_\Phi({\bf x}, \gamma {\bf x}^\prime)}
\def\xiugj{\xi^{u}_\Phi({\bf x},\gamma_j {\bf x}^\prime)}
\def\xiur{\xi^{u}_\Phi(r)}
\def\xiurj{\xi^{u}_\Phi(r_j)}
\def\xiaux{\xi^{u}_\Phi(\beta_*, r)}
\def\xicaux{\xi^{c}_\Phi(\beta_*,{\bf x}, {\bf x}^\prime)}
\def\dT{\frac{\Delta T}{T}}
\def\d{\delta}
\def\tauls{\tau_{\sc ls}}
\def\cobedmr{{\sc cobe--dmr\,}}
\def\chiH{{\chi_{\!{}_{\rm H}}}}
\def\ipert{\psi}
\def\ipertb{\psi_\beta}
\def\nn{\nonumber}

\newcommand{\pleg}[3]{ P^{#1}_{#2}(#3)}
\newcommand{\qleg}[3]{ Q^{#1}_{#2}(#3)}
\def\cosech{{\rm cosech}}
\def\shint{{\rm Shi}}
\def\chint{{\rm Chi}}
\def\sint{{\rm Si}}
\def\cint{{\rm Ci}}
\def\bqt{``}
\def\eqt{''}
\def\today{\ifcase\month\or
 January\or February\or March\or April\or May\or June\or
 July\or August\or September\or October\or November\or December\fi
 \space\number\day, \number\year}
\def\plotone#1{\centering \leavevmode
\epsfxsize= 0.6\columnwidth \epsfbox{#1}}

\def\plottwoside#1#2{\centering \leavevmode
\epsfxsize=.45\columnwidth \epsfbox{#1} \hfil
\epsfxsize=.45\columnwidth \epsfbox{#2}}

\begin{document}

\title {CMB Anisotropy in Compact Hyperbolic universes I: Computing
Correlation Functions } 
\author{J.Richard Bond, Dmitry Pogosyan and Tarun Souradeep} 
\address{Canadian Institute for Theoretical
Astrophysics,\\ University of Toronto, ON M5S 3H8, Canada}
\date{\today} \maketitle

\begin{abstract}
CMB anisotropy measurements have brought the issue of global topology
of the universe from the realm of theoretical possibility to within
the grasp of observations. The global topology of the universe
modifies the correlation properties of cosmic fields. In particular,
strong correlations are predicted in CMB anisotropy patterns on the
largest observable scales if the size of the Universe is comparable to
the distance to the CMB last scattering surface. We describe in detail
our completely general scheme using a {\em regularized method of
images} for calculating such correlation functions in models with
nontrivial topology, and apply it to the computationally challenging
compact hyperbolic spaces~\cite{us_texas,us_cwru}. Our procedure directly sums
over images within a specified radius, ideally many times the diameter
of the space, effectively treats more distant images in a continuous
approximation, and uses Cesaro resummation to further sharpen the
results. At all levels of approximation the symmetries of the space
are preserved in the correlation function. This new technique
eliminates the need for the difficult task of spatial eigenmode
decomposition on these spaces.  Although the eigenspectrum can be
obtained by this method if desired, at a given level of approximation
the correlation functions are more accurately determined. We use the
3-torus example to demonstrate that the method works very well. We
apply it to power spectrum as well as correlation function evaluations
in a number of compact hyperbolic (CH) spaces.  Application to the
computation of CMB anisotropy correlations on CH spaces, and the
observational constraints following from them, are given in a companion 
paper.
\end{abstract}

\section{Introduction}

The remarkable degree of isotropy of the cosmic microwave background
(CMB) points to homogeneous and isotropic Freidmann-Robertson-Walker
(FRW) models for the universe. The underlying Einstein's equations of
gravitation are purely local, completely unaffected by the global
topological structure of space-time. In fact, in the absence of
spatially inhomogeneous perturbations, a FRW model predicts an
isotropic CMB regardless of the global topology.

 Flat or `open' FRW models adequately describe the observed average
local properties of our Universe. Much of recent astrophysical data
suggest the cosmological density parameter in nonrelativistic matter,
$\Omega_m$, is subcritical~\cite{opencase}. The total density
parameter $\Omega_0$ includes relativistic particles and vacuum,
scalar field or cosmological constant, contributions, as well as
$\Omega_m$.  If a cosmological constant (or some other exotic smooth
matter component) does not compensate for the deficit from unity, this
would imply a hyperbolic spatial geometry (uniform negative
curvature), commonly referred to as an `open' universe in the
cosmological literature. The {\em simply connected} (topologically
trivial) hyperbolic 3-space $\hm$ and the flat Euclidean 3-space
${\cal E}^3$ are non-compact and have infinite volume.  There are
numerous theoretical motivations, however, to favor a spatially
compact universe~\cite{ell71,sok_shv74,lac_lum95,got80,cor9596}. To
reconcile a compact universe with a flat or hyperbolic geometry,
consideration of {\em multiply connected} (topologically non-trivial)
spaces is required. (Inhomogeneous simply connected models are another
way out; \eg a hyperbolic bubble could be embedded in a highly
inhomogeneous space and if the bubble is much larger than the Hubble
radius we would not know.) Compact hyperbolic spaces have been
recently used to construct cosmological models within the framework of
string theory~\cite{hor_mar98}. A compilation of recent articles
describing various aspects of current research in compact cosmologies
can be found in~\cite{stark98}.

The realization that the universe with the same local geometry has
many different choices of global topology is as old as modern
cosmology -- De Sitter was quick to point out that Einstein's closed
static universe model with spherical geometry ${\cal S}^3$ could
equally well correspond to a multiply connected {\em Elliptical}
universe model where the antipodal points of ${\cal S}^3$ are
topologically identified. The possibility of a multiply connected
universe has lingered on the fringes of cosmology largely as a
theoretical curiosity (\eg in \cite{ell71}), but was sometimes invoked
to explain puzzling cosmological observations (\eg as a possible
explanation for the isotropy of the CMB radiation \cite{got80}, and
for the (controversial) observations of periodic or discordant quasar
redshifts ~\cite{per_redshift}). There is a long history of attempts
to search for signatures of non-trivial global topology by identifying
ghost images of local galaxies, clusters or quasars at higher
redshifts~\cite{sok_shv74,lac_lum95,ghosts}.  The search for
signatures of global topology in the distribution of luminous matter
can probe the topology of the universe only on scales substantially smaller
than the apparent radius of the observable universe.  Another avenue
in the search for global topology of the universe is through the
effect on the power spectrum of cosmic density perturbation fields,
reflected in observables such as the distribution of matter in the
universe and the CMB anisotropy.

The observed large scale structure in the universe implies spatially
inhomogeneous primordial perturbations existed which gave rise to the
observed anisotropy of the CMB. The global topology of the universe
also modifies the local observable properties of the CMB anisotropy
on length scales up to a few times the horizon size. In compact
universe models, the finite spatial size usually precludes the
existence of primordial fluctuations with wavelengths above a
characteristic scale related to the size.  As a
result, the power in CMB anisotropy is suppressed on large angular
scales. Another consequence is the breaking of statistical isotropy in
characteristic patterns determined by the photon geodesic structure of
the manifold. One can search for such patterns statistically in CMB
anisotropy maps.  Full-sky CMB data, such as from the \cobedmr
experiment, can constrain the size of the universe and its topology to
the extent that such correlation patterns are absent in the
data~\cite{us_cwru,paperB}.

For Gaussian perturbations, the angular correlation function, $C(\hat
q,\hat q^\prime)$, of the CMB temperature fluctuations in two
directions $\hat q$ and $\hat q^\prime$ in the sky completely encodes
the CMB anisotropy predictions of a model.  For adiabatic perturbations, 
the dominant contribution
to the anisotropy in the CMB temperature measured with a wide-angle beam
($\theta_{\sc fwhm} \gta 2^\circ~ \Omega_0^{1/2}$) comes from the
cosmological metric perturbations through the Sachs-Wolfe effect. The
angular correlation function $C(\hat q,\hat
q^\prime)$ then depends on the spatial two point correlation function,
$\xi_\Phi \equiv \langle\Phi({\bf x},\tauls)\Phi({\bf
x^\prime},\tauls)\rangle $, of the gravitational potential, $\Phi$, on
the three-hypersurface of last scattering along the lines-of-sight,
${\hat q}$ and ${\hat q}^\prime$. Thus we need to learn how to compute
spatial correlation functions on compact spaces.

When the eigenfunctions of the Laplacian on the space are known, the
correlation function can be readily obtained via a mode sum. For this
to be tractable the known eigenfunctions would, preferably, be
expressed in a reasonably simple closed form. However, obtaining
closed form expressions for eigenfunctions may not be possible beyond
the simplest topologies. Some examples where explicit eigenfunctions
have been used include flat
models~\cite{star_angel,tor_refs,us_torus,janna} and a non-compact
hyperbolic space with a horn topology~\cite{sok_star75}.  In this
paper we provide a detailed description of our {\em regularized method
of images} for computing correlation functions, which does not require
any prior knowledge of the eigenfunctions on the compact
spaces~\cite{us_cwru,us_texas}. The method allows us to accurately
compute the correlation function on compact spaces where
eigenfunctions are not available. Most notorious in this respect are
compact spaces with uniform negative curvature, the {\em Compact
Hyperbolic} (CH) spaces, for which even numerical estimation of the
eigenfunctions is believed to be a challenging task (for recent
progress see \cite{eigen,eigen1,eigen2}).  A novel feature of our
scheme is the regularization procedure that we devise in order to
successfully implement the method of images for the power spectra of
cosmological perturbations expected from early universe physics. As an
additional bonus, this regularization scheme enhances the convergence
properties of the method, which proves to be very useful for tackling
CH spaces.

In $\S$\ref{sec_top}, we introduce compact spaces and briefly review
the aspects that are relevant for our work. The method of images is
derived in $\S$\ref{sec_moi}. As a simple illustrative example, we
apply the method of images to the case of a simple flat torus in
$\S$\ref{sec_tor}. The implementation of the method in CH spaces is
presented in $\S$\ref{sec_CH}. The gross properties of the power
spectrum in CH spaces that can be inferred from our computation is
discussed in $\S$\ref{ssec_CH_powspec}. We derive the properties of
the correlation function in CH space in $\S$\ref{sec_res}. In
$\S$\ref{sec_dis}, we discuss our results and point to our companion 
paper describing our detailed CMB anisotropy calculations~\cite{paperB}.

\section{Primer on Compact spaces}
\lbl{sec_top}

\subsection{Mathematical Preliminaries} \lbl{sec_mathprelim}

A compact cosmological model ${\cal M}$ is a {\em Quotient Space}
constructed by identifying points on the standard FRW spaces under the
action of a suitable~\footnote{ A Quotient space ${\cal M} = {\cal
M}^u/\Gamma$ consists of the equivalence classes of points of ${\cal
M}^u$ equivalent under $\Gamma$. ${\cal M}$ is a manifold when
$\Gamma$ is a subgroup of the isometry group, $G$ of ${\cal M}^u$
which acts freely (fixed point free: $\gamma {\bf x}={\bf x}
\Rightarrow \gamma = I$) and properly discontinously (every point
${\bf x}\in {\cal M}$ has a neighborhood $U_{\bf x}$ such that
$\gamma(U_{\bf x})\cap U_{\bf x}= \emptyset 
\, \forall\gamma\in\Gamma, ~\gamma \ne I$). The manifold ${\cal M}$ is compact if the
corresponding Dirichlet domain is compact.} discrete subgroup of
motions, $\Gamma$, of the full isometry group $G$, of the FRW space
(see~\cite{ell71,wol94}). The isometry group $G$ is the group of
motions which preserves the distances between points.  The infinite
FRW spatial hypersurface is the {\em universal cover}, ${\cal M}^u$,
tiled by copies of the compact space, ${\cal M}$. Cosmological
postulates of local homogeneity and isotropy restrict ${\cal M}^u$ to
be a space of constant curvature -- {\em Hyperbolic} $\hm$, with
negative curvature, {\em Spherical} ${\cal S}^3$, with positive
curvature, or flat {\em Euclidean} ${\cal E}^3$, with zero
curvature. The compact space for a given location of the observer is
represented as the {\em Dirichlet domain} with the observer at its
{\em basepoint}. Every point ${{\bf x}}$ of the compact space has an
image ${{\bf x}}_i = \gamma_i {{\bf x}}$ in each copy of the Dirichlet
domain on the universal cover, where $\gamma_i \in \Gamma$.  The
tiling of the universal cover with Dirichlet domains is a Voronoi
tessellation (a familiar concept in cosmology that has been used in
modeling the large scale structure in the universe), with the seeds
being the basepoint and its images. The Dirichlet domain represents
the compact space as a {\em convex polyhedron} with an even number of
faces, with congruent pairs of faces identified (glued) under
$\Gamma$. For more details, see \eg ~\cite{ell71,wol94,vin93,lac_lum95}.

More explicitly, the Dirichlet domain around the basepoint, ${\bf
  x}_0$, is the set of all points on the universal cover which are
closer (or equidistant) to ${\bf x}_0$ than to any of the images
$\gamma{\bf x}_0$ ($\gamma \in \Gamma$) of the basepoint. The
hyperplane that bisects the segment joining ${\bf x_0}$ to $\gamma{\bf
  x_0}$ divides ${\cal M}^u$ into two parts; let $H_\gamma$ denote
that half that contains the basepoint.  By definition, the Dirichlet
domain around ${\bf x}_0$ is ${\cal D}_{{\bf x}_0} = \bigcap_{\gamma}
H_\gamma$.  Thus, the Dirichlet domain is bounded by hyperplanes
bisecting the (geodesic) segments joining the basepoint ${\bf x}_0$
to a set of adjacent images. (The corresponding set of motions is
called the set of {\em adjacency transformations}.) The faces of the
polyhedron are identified pairwise; the one formed by the hyperplane
bisecting the segment joining ${\bf x_0}$ to $\gamma{\bf x_0}$ is
identified with the corresponding one bisecting the segment joining
${\bf x_0}$ to $\gamma^{-1}{\bf x_0}$.  The adjacency transformations
generate the group $\Gamma$ and hence are also known as the {\em face
  generators}.

In the context of cosmology, the Dirichlet domain constructed around
the observer represents the universe as `seen' by the observer. (In
our papers, we will often loosely use the same notation ${\cal M}$ to
refer to the compact space as well as one of its Dirichlet domain
representations whose basepoint is clear from the context.)  It proves
useful then to define the {\em outradius} $R_>$ and the {\em inradius}
$R_<$ of the Dirichlet domain~\cite{sok_shv74}: the radius of the
smallest sphere around the observer which encloses the Dirichlet
domain and that of the largest sphere around the observer which can be
enclosed within the Dirichlet domain, respectively . The ratio
$R_>/R_<$ is a good indicator of the shape of the Dirichlet domain. 

In cosmology, distances are inferred from the cosmological redshift
which is related to the light travel time. Consequently, in compact
spaces, and, more generally, in multiply connected spaces, light from the
same source seems to arrive from points at different locations -- the
source and its images on the universal cover.  Any source inferred to
be at a distance greater than $R_>$ from the observer is bound to be
the image of a point which is physically closer. On the other hand, a
source closer than $R_<$ is definitely at its true physical distance.
Note that $R_>$ and $R_<$ are specific to the location of the observer
within the compact space since the Dirichlet domain around different
observers can, in general, vary.  An observer (and Dirichlet domain)
independent linear measure of the size of the compact space is given
by the {\em diameter} of the space, $d_{\!\cal M} \equiv
\sup_{x,y\in{\cal M}}d(x,y)$, \ie the maximum separation between two
points in the compact space. 

The isometry group $G$ defining the global symmetries of ${\cal M}$ is
the {\em centralizer} of $\Gamma$ in the isometry group $G^u$ of its
universal cover, ${\cal M}^u$; i.e., $G = \{ g\in G^u | g\gamma=\gamma g
\, \forall \gamma\in\Gamma\}$. In general, ${\cal M}$ respects less
symmetries than ${\cal M}^u$; consequently, $G$ is of lower dimension
than $G^u$. ${\cal M}$ is (globally) homogeneous if and only if $G$ is
{\em transitive} on ${\cal M}$: \ie for any two points ${\bf x},{\bf
y} \in {\cal M}$, there exists a $g\in G$ such that ${\bf y} = g
{\bf x}$ (an equivalent statement is that a compact space is globally
homogeneous {\em if and only if} every element of $\gamma\in\Gamma$
is a {\em Clifford translation}, \ie the {\em displacement function},
$d_\gamma (x) \equiv d({\bf x}, \gamma{\bf x})$, is independent of 
${\bf x}$ for all points ${\bf x}\in {\cal M}^u$). Space is isotropic
(around a point, or observer, ${\bf x_o}$) if $G$ contains a subgroup of rotations
around ${\bf x_o}$. The only example of a multiply
connected compact universe which retains all the symmetries of its
universal cover is the Elliptic space (${\cal S}^3/Z_2$) with
spherical geometry; the simple flat torus is anisotropic, and
all others (including the entire class of CH manifolds) break global
homogeneity as well. As the result, the two-point correlation functions
$\xi({\bf x,y})$
in CH spaces depend separately on both points ${\bf x}$ and ${\bf y}$,
and not only on the distance $d({\bf x,y,})$ as in the familiar FRW spaces. 

\subsection{Compact Euclidean spaces}
\label{ssec_top_fl}

The compact spaces with Euclidean geometry (zero curvature) have been
completely classified. In three dimensions, there are known to be six
possible topologies that lead to orientable
spaces~\cite{ell71,wol94,vin93}: the simple flat torus where ${\cal
E}^3$ is identified under a discrete group of translations; three other
flat tori where the identification is under a {\it screw motion}, \ie
translation accompanied by a rotation about the direction of the
translation, namely, rotation by $\pi$ or $\pi/2$ in one component of
the translations, and rotation by $\pi$ in all three directions; and
finally, two topologies made by identifying the planar hexagonal
lattice under a screw motion perpendicular to the plane with rotation
of $\pi/3$ and $2\pi/3$, respectively.

\subsection{Compact Hyperbolic spaces}
\label{ssec_top_CH}

For cosmological CH models, ${\cal M}^u \equiv \hm$, the three
dimensional hyperbolic (uniform negative curvature) manifold with line
element
\begin{equation}
d s^2 = d\chi^2 + \sinh^2\chi\,(d\theta^2 + \sin^2\theta d\phi^2)\, ,
\label{h3metric}
\end{equation}
where $\chi = (\tau_0 -\tau)/d_c$ is the affine distance, $\tau$ is
the conformal time and $d_c = c H_0^{-1}/\sqrt{1 -\Omega_0}$ is the
curvature radius set by the present cosmological density
parameter $\Omega_0$ and the Hubble constant $H_0$.\footnote{Unless
explicitly indicated, all distances and times in non-flat geometry are
in units of the curvature scale $d_c$.} For a universe with hyperbolic
geometry, $0< \Omega_0 < 1$; thus the size of $d_c$ ranges
from $c H_0^{-1}$ as $\Omega_0\to0$ to infinity as $\Omega_0 \to 1$.

$\hm$ can be viewed as a hyperbolic section embedded in four
dimensional flat Lorentzian (Minkowski) space, by representing each
point on $\hm$ as a unit four vector, ${\bf X}\equiv(\zeta({\bf x}),
{\bf x})$, normalized by $d_c$ in Minkowski space ($\zeta^2 -|{\bf
x}|^2 = d_c^2$). The distance between two points on $\hm$ is given
by the dot product of the corresponding four vectors.  The isometry group
of $\hm$ is then the group of rotations in the four space, the proper
Lorentz group $SO(3,1)$.  A CH manifold is completely described by a
discrete subgroup, $\Gamma$, of the proper Lorentz group $SO(3,1)$.

There are two remarkable features of tessellating $\hm$ under a
discrete group of motions that are absent in the flat geometry.
Firstly, whereas in flat geometry all finite volume Quotient Spaces
obtained are necessarily compact, one can tessellate $\hm$ with tiles
of finite volume which are non-compact, giving rise to a class of
non-compact finite-volume hyperbolic universe models.
Typically, these spaces have cusp-like extensions to infinity.
Secondly, a given CH topological structure fits only for a specific
volume of the space. This is in contrast to flat compact spaces,
in which the same topological structure can be
imposed on any scales. In particular, all simple flat tori with different
identification lengths are {\em homeomorphic} to each other, \ie one
can be obtained from the other by a continuous mapping/deformation,
but are {\em not isometric} to each other, \ie the distance between the
mapped points are not the same. 
Two hyperbolic spaces of finite volume which are homeomorphic
are necessarily isometric (in three dimensions and
above). This result, known in mathematics
literature as the {\em strong rigidity theorem}, can be attributed to
the existence of the intrinsic length scale on $\hm$ -- the curvature
radius, $d_c$.  It not difficult to convince oneself that the
Dirichlet domain has to have a fixed size relative to $d_c$ in order
to tile $\hm$ -- if one were to scale the polygon size (relative to
$d_c$) then the scaled tiles will no longer fit together at the
vertices.

Thus, a CH manifold, ${\cal M}$, is characterized by a dimensionless
number, ${\cal V}_{\!\cal M}\equiv V_{\!\cal M}/d_c^3$, where
$V_{\!\cal M}$ is the volume of the space and $d_c$ is the curvature
radius \cite{Thur7984}. There are a countably infinite number of CH
manifolds, and no upper bound on ${\cal V}_{\!\cal M}$. The
theoretical lower bound stands at ${\cal V}_{\!\cal M} \ge
5/(2\sqrt{3}) {\rm Arcsinh}^2 \left(\sqrt{3}/5\right)\approx
0.167$~\cite{gab_mey96}. The smallest CH manifold discovered so far
has ${\cal V}_{\!\cal M} =0.94$ and it has been conjectured that this
is, in fact, the smallest possible~\cite{smallestCH1,smallestCH2}.
The physical volume of the CH universe with a given topology, \ie a
fixed value of $V_{\!\cal M}/d_c^3$, is then set by $d_c$ and is thus 
related to cosmological parameters. The Geometry Center at the
University of Minnesota has a large census of CH manifolds in its
public domain. This census was created using the SnapPea computer
software which is also freely available at the website~\cite{Minn}.
The {\em Minnesota census} lists several thousands of these manifolds
with ${\cal V}_{\!\cal M}$ up to $\sim 7$ and the SnapPea software can
be used to obtain various characteristic properties of these CH
manifolds such as the ones listed in Table~\ref{tab:CHprop} and also the
generators of the discrete group motion $\Gamma$ that we need for our
computational method.

The sequences of CH manifolds are closed well-ordered sets of order
type $\omega^\omega$, \ie are arranged in a countably infinite number
of countably infinite sequences (in increasing volume), with each
sequence having a non-compact finite volume hyperbolic space as its
limit. These limit spaces of finite volume have cusps extending to
infinity. The cusps are homeomorphic to $T^2\times R^+$, the product
of a $2$-torus and the positive real line. The sequence of CH spaces
arises through {\em Dehn filling}, a procedure which truncates the
cusps of the limit space along a closed curve on $T^2$ (with allowed
pair of winding numbers) and glues in a solid torus (by identifying
the closed curve along a meridian of the solid torus).  The
nomenclature of CH spaces in the Minnesota census, which we follow in
this work, identifies a sequence by the limit space and specifies a CH
space belonging to it by two (coprime) integer indices in parenthesis
correspond to the winding numbers, $m$ and $n$, which characterize the
closed curve on the torus; \eg m004(-5,1) refers to the CH manifold
obtained by Dehn-filling the single cusp of the non-compact space m004
by a torus glued along a closed curve with winding numbers $(-5,1)$.
The limit non-compact space is achieved as a limit of large values of
the winding numbers in the corresponding sequence of CH manifolds.
From the observational perspective, the CH spaces with high winding
numbers should be practically indistinguishable from the limit
non-compact space.  The space with winding numbers $(m,n)$ is
equivalent to the space $(-m, -n)$ in the same sequence. Not all
integer values of $m$ and $n$ lead to a CH space, there are bounded
forbidden gap regions in the $m$--$n$ lattice. One should note that
the volume does not uniquely characterize a space; there can be a
finite number of distinct CH spaces with the same volume (besides the
mirror image pairs for chiral CH spaces such as m004(-5, 1) \& m004(5, 1)).
Also, the nomenclature described above is not unique: the same CH
space may belong to two different sequences, \eg the space m003(-2,3)
is equivalent to m004(5,1). (See \cite{smallestCH2} for details.)

All CH hyperbolic spaces are necessarily globally inhomogeneous since
the only element of $SO(3,1)$ which is a Clifford translation is the
identity~\cite{wol94}. A short proof follows.  Assume that $g \in
SO(3,1)$ is a Clifford translation. Then $d({\bf x}_1,g {\bf x}_1) =
d({\bf x}_2,g {\bf x}_2)$ for any two points ${\bf x}_1$ and ${\bf
x}_2$ on $\hm$. Since $SO(3,1)$ is the isometry group of the
homogeneous space $\hm$, there exists a Lorentz transformation 
$\Lambda \in SO(3,1)$ such that ${\bf x}_2 = \Lambda {\bf
x}_1$. Consequently, $d({\bf x}_1,g {\bf x}_1)= d(\Lambda{\bf x}_1,g
\Lambda{\bf x}_1)$, implying that $g = \Lambda^{-1} g \Lambda$\,
$\forall \Lambda \in SO(3,1)$. Since the Lorentz group is {\em
non-Abelian} (\ie actions of the group elements do not commute), $g$
must be the identity.

\begin{figure}
\plottwoside{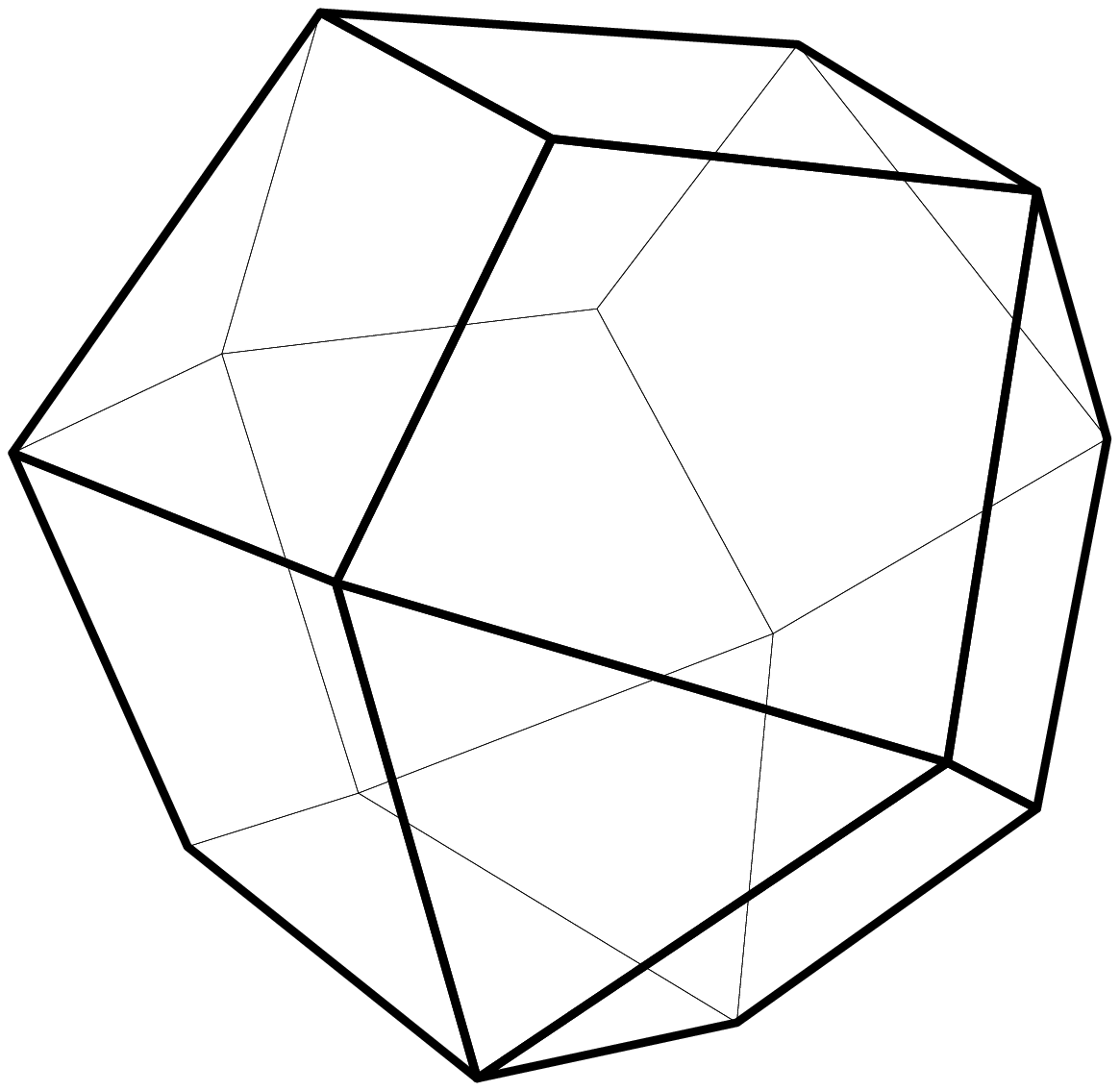}{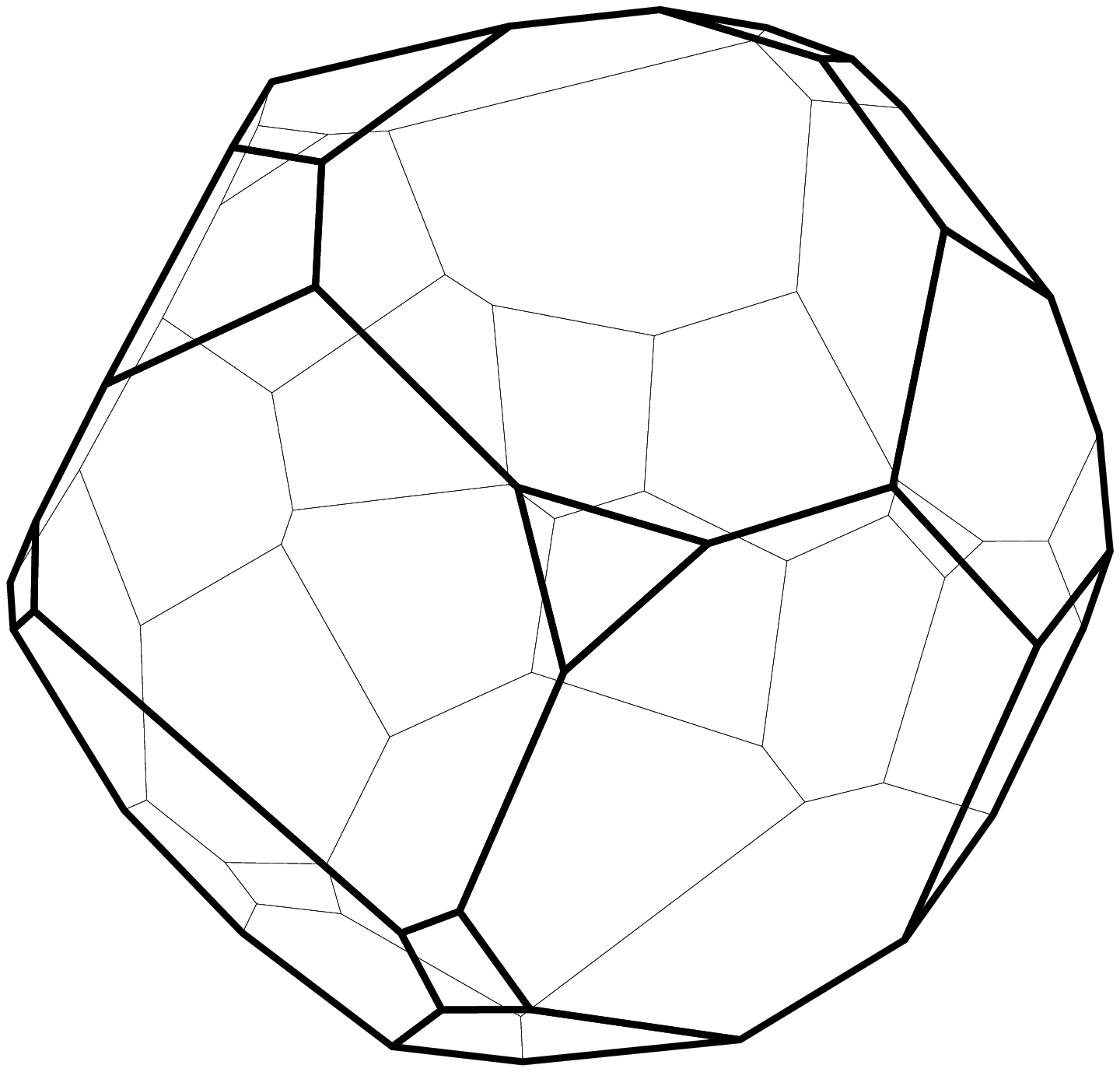}
\caption{The Dirichlet domains of two of the compact hyperbolic (CH)
spaces that we have studied are shown. On the left is a small CH
space, m004(-5,1), and on the right is a large CH space, v3543(2,3).}
\label{fig:dirdoms}
\end{figure}
\begin{table}
\caption{The characteristics of some of the compact hyperbolic (CH)
manifolds that we have studied. The nomenclature of the spaces
conforms to the Minnesota census. The volume and diameter (expressed
in the units of curvature radius) are topological invariants which
relate to the density of states in those spaces.  The specifications
have been obtained from the Minnesota census using SnapPea. The exception
is the diameter which is our estimate using a simple random
sampling algorithm.  However, in the context of cosmology, the
observer specific characterization of the Dirichlet domain (around the
observer) proves useful. The m003 models correspond to spaces obtained
by different Dehn fillings on a (non-compact) cusped manifold,
m003. The CH space m003(-3,1) has the smallest volume in the m003
series and is also currently the smallest CH space known. The
m004(-5,1) has the smallest volume in the m004 series. The v3543(2,3)
space is a relatively large one.}
\begin{tabular}{lccccccc}
\lbl{tab:CHprop}

Properties &m003(-3,1)&m003(-4,3)&m003(-4,1)&m003(2,3)& m003(-5,4)
&m004(-5,1)& v3543(2,3)\\
\tableline
Topological invariants:& & & & & & &  \\
Volume: ${\cal V}_{\!\cal M}$ &0.94 &1.26 &1.42&1.54&1.59&0.98&6.45\\
diameter: $d_{\!\cal M}$&0.84&1.01 &1.10&1.16&1.21& 0.86& 1.90 \\ 
First Homology group &$Z_5\oplus Z_5$ &$Z_5\oplus Z_5$ &$Z_{35}$ 
& $Z_{35}$ &$Z_{30}$ & $Z_5$ & $Z_{93}$  \\ 
\tableline
Dirichlet domain specific: &  & & & & &  &  \\
outradius: $R_>/d_c$ &0.75&0.84&1.07&0.83& 0.94&0.75 &1.33\\
inradius: $R_</d_c$ &0.52&0.55&0.54&0.59&0.57 &0.53 & 0.89\\
        No. of faces   &18&22&24&26 &28&16&38\\
        No. of vertices&26&40&44&48&52&16&72\\
\end{tabular}
\end{table}

\section{Method of Images}
\lbl{sec_moi}
\subsection{The correlation function}

The correlation function $\xi_\Phi({\bf x},{\bf x}^\prime)$ of a
scalar field, $\Phi$, can be expressed formally in terms of the
orthonormal set of eigenfunctions $\Psi_i$ of the Laplace operator,
$\nabla^2$, on the hypersurface (with positive eigenvalues $k_i^2 \ge
0$), as \cite{chav84}
\begin{eqnarray} 
\xi_\Phi({\bf x}, {\bf x}^\prime) = \sum_i P_\Phi(k_i)
 \sum_{j=1}^{m_i} \Psi_{ij}({\bf x})\Psi_{ij}^*({\bf x}^\prime) \, , \
 {\rm where}\ (\nabla^2 + k_i^2)\Psi_{ij} = 0\,. \lbl{xi1}
\end{eqnarray}
The spectrum of the Laplacian on a compact space (thus with closed
boundary conditions) is a discrete ordered set of eigenvalues
$\{k^2_i\}$ ($k_0^2 = 0$ and $k_i^2 < k^2_{i+1}$) with 
multiplicities $m_i$.  The function $P_\Phi(k_i)$ describes the {\it
rms} amplitude of the eigenmode expansion of the field $\Phi$, 
determined in the context of cosmology by the physical mechanism
responsible for the generation of $\Phi$.

Except in simple cases, neither the spectrum of the Laplacian
$\{k^2_i\}$ nor the eigenfunctions $\Psi_{ij}^c({\bf x})$ are known
for compact manifolds, so eq.(\ref{xi1}) cannot be used to calculate
the correlation function $\xic$ on a compact manifold ${\cal M}$
directly.  In contrast, the universal cover ${\cal M}^u$ for the
compact manifold is usually simple enough (\eg $\hm$, ${\cal S}^3$ or
${\cal E}^3$) that the eigenfunctions $\Psi^u_j(k,{\bf x})$ are known
and the correlation functions $\xiu$ are easily computable.
We consider $\hm$ and ${\cal E}^3$ geometries where the spectrum of
eigenvalues on the universal cover ${\cal M}^u$ is continuous, 
reflected in the notation $\Psi^u_j(k,{\bf x})$ replacing the discrete
index $i$ with a functional dependence on $k$.

The {\em regularized method of images} we developed in \cite{us_texas}
allows computation of the correlation function $\xic$ on a compact
(more generally, a multiply connected) manifold ${\cal M}= {\cal
  M}^u/\Gamma$ from the correlation function, $\xiu$, on ${\cal M}^u$.
We now give the explicit derivation of the relation between $\xic$ and
$\xiu$, calculated with the same form of the power spectrum
$P_\Phi(k)$. The method solely relies on the knowledge of the action
of elements of the discrete group, $\Gamma$, and requires no
information regarding the eigenvalues and eigenmodes of the Laplacian
on ${\cal M}$.  The expansion (\ref{xi1}) is equivalent to the system
of integral equations on the correlation functions, $\xi^c_\Phi$ and
$\xi^u_\Phi$,

\begin{mathletters}
\begin{eqnarray}
\int_{{\cal M}} d{\bf x^\prime}~\xic~\Psi^c_{ij}({\bf x}^\prime) =
P_\Phi(k_i)\Psi^c_{ij}({\bf x})\,,\lbl{eigeqn3}\\ \int_{{\cal M}^u}
d{\bf x^\prime}~\xiu~\Psi^u_{j}({k,\bf x^\prime}) =
P_\Phi(k)\Psi^u_{j} (k,{\bf x})\,.  \lbl{eigeqn4}
\end{eqnarray}
\end{mathletters}
Eq.(\ref{eigeqn4}) is a consequence of the homogeneity of ${\cal M}^u$
through a theorem~\cite{chav84} which states that the eigenfunctions
of the Laplacian are also the eigenfunctions of the integral operator
corresponding to any two-point function which is {\em point-pair
invariant}, \ie depends only the distance between the points. The
orthonormality of the eigenfunctions leads to the expansion
(\ref{eigeqn3}) for $\xi^c_\Phi$.

Every eigenfunction $\Psi^c_{ij}({\bf x})$ on ${\cal M}$ is also an
eigenfunction of the Laplacian on the universal cover ${\cal M}^u$
with eigenvalue $k_i^2$, hence is a linear combination of degenerate
eigenfunctions $\Psi^u_{j}(k_i,{\bf x})$ on ${\cal M}^u$ with
eigenvalue $k_i^2$, \ie $\Psi^c_{ij}({\bf x})=\sum_{j^\prime}
a_{i,jj^{\prime}} \Psi^u_{j^{\prime}}(k_i,{\bf x})$. Thus a subset of
equations (\ref{eigeqn4}) can be written as
\begin{eqnarray}
\int_{{\cal M}^u} d{\bf x^\prime}~\xiu~\Psi^c_{ij}({\bf x}^\prime) &=&
P_\Phi(k_i)\Psi^c_{ij}({\bf x})\\ \lbl{eigeqn2}
&=& \int_{{\cal M}} d{\bf x^\prime}~\xic~\Psi^c_{ij}({\bf x}^\prime)\,.
 \nonumber
\end{eqnarray}
Using automorphism of $\Psi^c_{ij} $ with respect to $\Gamma$,
$\Psi^c_{ij}(\gamma{\bf x})= \Psi^c_{ij}({\bf x})\, \forall
\gamma\in\Gamma$, and the fact that ${\cal M}$ tessellates ${\cal M}^u$,
\begin{mathletters} 
\begin{eqnarray}
\int_{{\cal M}} d{\bf x^\prime}~\xic~\Psi^c_{ij}({\bf x}^\prime) &=&
\int_{{\cal M}^u} d{\bf x^\prime}~\xiu \Psi^c_{ij}({\bf x^\prime}) =
\sum_{\gamma\in\Gamma}\int_{\cal M} d{\bf x^\prime}~ \xiug~
\Psi^c_{ij}({\bf x^\prime}) \lbl{kernelsuma} \\ &=& \int_{\cal M}
d{\bf x^\prime}~\left[~\widetilde{\sum_{\gamma\in\Gamma}} \xiug ~\right]
\Psi^c_{ij}({\bf x^\prime}), \lbl{kernelsumb}
\end{eqnarray}
\end{mathletters} 
where $\widetilde{\sum} $ denotes a possible need for regularization
at the last step when the order of integration and summation is
reversed. Since eq.(\ref{kernelsumb}) is satisfied for all
$\Psi^c_{ij}$, comparison with eq.~(\ref{eigeqn3}) implies that
\begin{equation}
\xic = \widetilde{\sum_{\gamma\in\Gamma}} \xiug\,.
\label{eq:mainim}
\end{equation}
This is the main equation of our {\em method of images}, which
expresses the correlation function on a compact space (and more
generally, any non-simply connected space) as a sum over the
correlation function on its universal cover calculated between ${\bf
x}$ and the images $\gamma{\bf x}^\prime$ ($\gamma\in\Gamma$) of ${\bf
x^\prime}$.

Regularization in eq.(\ref{kernelsumb}) is required when the
correlation function on the universal cover does not have a compact
support, $\int_{{\cal M}^u} \xiu d{\bf x^\prime} =\infty $. Let us
consider a new two-point function on ${\cal M}^u$ defined by
\begin{equation}
\tilde{\xi}^u_{\Phi}({\bf x,x^\prime}) \stackrel{\mathrm{def}}{=} \xiu
- \frac{1}{V_{\!\cal M}}\int_{\gamma{\cal M}} d{\bf x}^{\prime\prime}~
\xi^u_{\Phi} ({\bf x},{\bf x^{\prime\prime}}), 
\label{eq:moi0}
\end{equation}
for $\gamma$ such that ${\bf x}^\prime$ lies in $\gamma {\cal M}$.
Replacing $\xiu$ by its regularized version $\tilde\xi^u_{\Phi}({\bf
x,x^\prime})$ in eq.(\ref{kernelsuma}) gives the same integral
equation on $\xic$ as the original one for every eigenfunction
$\Psi^c_{ij}$ with the exception of that for the zero mode
$k^2=0$. For the zero mode, $\Psi^c_{0}=\mathrm{const}$, for which the
right-hand side of eq.~(\ref{eq:moi0}) is now zero~\footnote{The
volume integral of the Laplace equation on a compact manifold ${\cal
M}$ (thus with closed boundary) gives $k_i^2 \int _{\cal M} d{\bf
x^\prime} \Psi^c_{ij}({\bf x})=0\, ~\forall i,j$.}.
Also, $\int_{{\cal M}^u} \tilde\xi^u_{\Phi}({\bf
x,x^\prime}) d{\bf x^\prime} = 0$.  Thus, the regularized sum over
images is
\begin{eqnarray}
\xic = \sum_{\gamma\in\Gamma}\tilde\xi^u_{\Phi}({\bf x},\gamma{\bf
x^\prime}) = \sum_{\gamma\in\Gamma} \xiug -\frac{1}{V_{\!\cal
M}}\int_{{\cal M}^u}d{\bf x^{\prime\prime}}~\xi_{\Phi}^u({\bf x}, {\bf
x^{\prime\prime}})\,. \lbl{moi1}
\end{eqnarray}
One can verify that $\xic$ is {\em biautomorphic} with respect to
$\Gamma$, \ie $\xi_\Phi^c({\bf x}, {\bf x}^\prime) =
\xi_\Phi^c(\gamma_1{\bf x}, \gamma_2{\bf x}^\prime) ~\forall
\gamma_1,\gamma_2 \in \Gamma$.  In addition, $\xi_\Phi^c$ is smooth
and symmetric, $\xi_\Phi^c({\bf x}, {\bf x}^\prime)= \xi_\Phi^c({\bf
x}^\prime, {\bf x})$, since $\xi^u_{\Phi}$ is smooth and point-pair
invariant. With these conditions satisfied, the correlation function
then qualifies to be the kernel of an integral operator on square
integrable functions on ${\cal M}$.

The sum in eq.~(\ref{moi1}) can be obtained analytically in a
closed form only in a few cases, \eg the simple flat torus (see
section ~\ref{sec_tor}).  For a numerical implementation of the sum
over images, it is useful to present the formal expression
(\ref{moi1}) as the limit of a sequence of partial sums
\begin{equation}
\xic = \lim_{N \to \infty}
\sum_{i=0}^N \left[ \xi_{\Phi}^u({\bf x},\gamma_i{\bf x^{\prime}})
-\frac{1}{V_{\!\cal M}}\int_{{\cal M}}d{\bf x^{\prime\prime}}~
\xi_{\Phi}^u({\bf x},\gamma_i{\bf x^{\prime\prime}})\right]\,,
\lbl{eq:series}
\end{equation}
with the discrete motions $\gamma_i$ sorted in increasing separation,
$d({\bf x},\gamma_i {\bf x^{\prime}}) \le d({\bf x},\gamma_{i+1} {\bf
x^{\prime}})$.  In practice, the summation over images is carried out
over a sufficiently large but finite number of images, from which the
limit $N \to \infty$ is estimated.  Accuracy is enhanced if the
$\gamma_i$ used correspond to a tessellation of ${\cal M}^u$ with the
basepoint of the Dirichlet domain shifted to ${\bf x}$.

A more effective and simpler limiting procedure for implementing the
regularized method of images is to explicitly sum images up to a
radius $r_*$ and regularize by subtracting the integral $\xiur$ over a
spherical ball of radius $r_*$. This further eliminates the need to
know the precise shape of Dirichlet domains (and integrate
$\xi_\Phi^u$ over potentially complicated shapes). We use this
limiting procedure when dealing with CH spaces (see
eq. (\ref{moi_final}) in $\S$\ref{ssec_CH_numer}). The example of the
simple flat torus discussed in $\S$\ref{sec_tor} shows that the radial
limiting procedure does better in eliminating unwanted low $k$ power
than eq.~(\ref{eq:series}). On the other hand, the radial limiting
procedure cannot be applied unambiguously if the number of images used
is small whereas eq.(\ref{eq:series}) holds at all values of $N$.

As we see, the need for regularization is strictly dictated by the
form of $\xi^u$, which encodes how modes are excited, and is not
specific to CH spaces. Some authors~\cite{eprolif} have incorrectly
attributed the need for the regularization that we invoke in
~\cite{us_texas,us_cwru} to the exponential proliferation of periodic
orbits or the chaotic nature of classical trajectories in CH
spaces. Regularization is required in flat compact spaces too (see
$\S$\ref{sec_tor}) if the $\xi^u$ mode expansion contains the zero mode
$k^2=0$.  On the other hand, if $\xi^u$ has compact support there is
no need for regularization even in CH space; this, in fact, holds
under weaker conditions.

The counter term in eq.({\ref{eq:series}) significantly improves the
convergence of the sum over images, even if in the limit $N\to\infty$
the term is zero ( \ie $\int_{{\cal M}^u} \xiu d{\bf x^\prime} = 0 $)
and regularization is formally not required. Indeed, in this case the
$N$th partial term of eq.({\ref{eq:series}) is equal to
\begin{equation}
\xi_\Phi^c({\bf x,x^\prime})_{\vert_N} = \sum_{i=0}^N \xi_{\Phi}^u({\bf
x},\gamma_i{\bf x^{\prime}}) +\frac{1}{V_{\!\cal M}}\int_{\bar{\cal
M}_N}d{\bf x^{\prime\prime}}~ \xi_{\Phi}^u({\bf x,x^{\prime\prime}}) \,.
\label{eq:Nth}
\end{equation}
where $\bar{\cal M}_N = {\cal M}^u - \cup_{i=0}^N \{\gamma_i {\cal
M}\}$ is the complement relative to ${\cal M}^u$ of the domains from
which the image contribution has been explicitly summed.  Thus,
$\xi_\Phi^c({\bf x,x^\prime})_{\vert_N}$ corresponds to the
approximation where the first $N$ images up to the distance $d \le
d({\bf x},\gamma_N {\bf x^\prime})$ are summed explicitly and the
contribution of the rest of the images is estimated as the
integral. The latter estimation is quite natural, since the sum over
densely packed distant images is similar to a Monte-Carlo expression
for the integral.

Finally, we note that the regularization term is certainly not
unique. In fact, instead of starting with a regularized correlation
function on the universal cover as in eq.~(\ref{eq:moi0}), we could
have started with a regularized scalar field, $\tilde\Phi = \Phi
-\int_{\cal M} dV \Phi /V_{\cal M}$. This gives rises to different
regularizing counterterms that might possibly be more effective but
are also more complicated. In the same spirit of viewing the field as
the primary starting point, the sum-over-images representation for
$\xic$ can be viewed as a double sum $\sum_\gamma \sum_{\gamma^\prime}
\xi^u_\Phi(\gamma {\bf x}, \gamma^\prime{\bf x}^\prime)$. In general
this is computationally more expensive. However, for the case of the
simple flat torus, the double sum can be expressed as a single
summation over appropriately weighted image contributions and has a
significantly faster convergence.

\subsection{The Power Spectrum}
\lbl{ssec_powspec}

An illuminating way to present the power spectrum of fluctuations in
compact spaces is to separate the {\em density of states} (the
spectrum of eigenvalues weighted with multiplicity), solely determined
by the geometry and topology of the space, from the {\em occupation
number} ({\it rms} amplitudes of the eigenmodes) determined by the
physics invoked to excite the available modes. This allows a
discussion of the effect of non-trivial topology independent of the
generation mechanism and the resultant statistical nature and spectrum
of the initial perturbations.
 
We define the power spectrum ${\cal P}_\Phi^c(k,{\bf x})$ of the
scalar field $\Phi$ to be the function which describes the
contribution to the variance of the field $\sigma_{\Phi}^2({\bf x})$
from the modes in a logarithmic interval of eigenvalues $d\ln k$.  The
variance is given by the correlation function at zero lag, thus our
definition satisfies $\xico=\int_0^{\infty} d \ln (k) {\cal P}_\Phi^c
(k,{\bf x}) $. The power spectrum depends both on the eigenvalue
spectrum of the Laplacian, described through a collapsed two point
function, $n^c(k, {\bf x},{\bf x})$, as well as on the {\it rms}
amplitudes $P_{\Phi}(k)$ of the eigenmode expansion of the field on
the universal cover:
\begin{mathletters}
\begin{eqnarray}
{\cal P}_\Phi^c (k,{\bf x}) \,d \ln (k) =  P_\Phi(k)\, n^c(k,{\bf
x},{\bf x})\, d\ln(k)\,, \\ n^c(k,{\bf x},{\bf x}) = k \sum_i \,
\delta(k-k_i) S_{k_i}({\bf x}), \\ S_{k_i}({\bf x}) = \sum_j
{\left|\Psi^c_{ij}({\bf x})\right|}^2\,.
\label{cct_powspec}
\end {eqnarray}
\end{mathletters}
The function $n^c(k, {\bf x},{\bf x})$ itself may be interpreted as
the local density of states (per $d \ln k$) {\it per unit volume} at
the point ${\bf x}$.  The notation $n^c(k,{\bf x},{\bf x})$ explicitly
retains the intimate connection to a two-point kernel.  Indeed,
$S_{k_i}({\bf x})$ is related to the imaginary part of the Green
function~\cite{sok_star75}
\begin{equation}
S_{k_i}({\bf x}) = -\frac{2 k_i}{\pi}~{\mathrm Im}[G_{k_i}({\bf
x},{\bf x})] \, . \lbl{cct_powspec2}
\end{equation}
The dependence on the position is a manifestation of the global
inhomogeneity of the space.  In the case of a globally homogeneous
space, $G_{k_i}({\bf x},{\bf x}^\prime)=G_{k_i}({\bf x}-{\bf
x}^\prime)$, implying that $S_{k_i}({\bf x})$ and $n(k, {\bf x})$ are
position independent.  We shall denote mean density of
states by $n^c(k)$, just omitting spatial dependence.
Integrating $S_{k_i}({\bf
x})$ over the volume of the manifold ${\cal M}$ gives the multiplicity
of the $i$th eigenvalue $\int _{\cal M} S_{k_i}({\bf x}) d{\bf
x}=m_i$, hence
\begin{equation}
n^c(k) = \frac{1}{{\cal V}_{\cal M}} 
\int _{\cal M} n^c(k,{\bf x},{\bf x})\; d{\bf x} = 
\frac{1}{{\cal V}_{\cal M}} \sum_i k \, \delta(k-k_i) m_i\, .
\label{eq:nck}
\end {equation}
Thus we factor the power spectrum in a compact space into a
{\it rms} amplitude of modes and the density of states: 
\begin{equation}
 {\cal P}_\Phi^c (k) = \frac{1}{{\cal V}_{\cal M}} \int _{\cal M} {\cal
P}_\Phi^c (k,{\bf x})\; d{\bf x} = P_\Phi(k)\, n^c(k)\,.
\end {equation}

In the universal covering space ${\cal M}^u$ with a continuous spectrum
of eigenstates, the density of states kernel is given by
\begin{equation}
 n^u(k,{\bf x},{\bf x}^\prime) = \frac {k^3}{(2 \pi)^3} \sum_j
\Psi^u_j(k,{\bf x}) \Psi^{u*}_j(k,{\bf x}^\prime)\,.
\label{eq:nuk}
\end{equation}
The homogeneity of ${\cal M}^u$ implies that $ n^u(k,{\bf
x},{\bf x})= n^u(k)=k^3/(2\pi^2)$ is independent of ${\bf x}$.
Akin to the correlation function, the method of images can be used
to establish
the connection between the local density of states kernel on the
compact manifold ${\cal M}^c$ and on its universal covering space
${\cal M}^u$:
\begin{equation}
 n^c(k, {\bf x}, {\bf x})=\sum_{\gamma} n^u(k,{\bf x},\gamma{\bf x})\, . 
\label{eq:nckx}
\end{equation}

The method of images applied to the density of states, together with
the connection to the trace of the Green function from
eq.(\ref{cct_powspec2}), is what is essentially embodied in the
celebrated Selberg trace formula~\cite{chav84}. We emphasize that,
although the basic ideas are similar, the computation of the
correlation function and the density of states $ n^c(k)$ in a compact
space are distinct problems.  Getting the density of states is
computationally more challenging since the aim is to recover a
singular function (string of delta functions) by superposing smooth
functions. For computing correlations of the CMB anisotropy in compact
spaces, we only need to apply the method of images to the correlation
function. Also, the pairwise correlation function at distinct points
${\bf x} \ne {\bf x^\prime}$, calculated by the method
of images at any level of approximation,
satisfies exactly the periodicity of the space,
which the density of states does only if determined with absolute precision.

\section{Flat torus model: a simple example}
\lbl{sec_tor}

The simple flat torus model, $T^3$, is the compactification of the
3-dimensional Euclidean space which identifies points under a discrete
set of translations, ${\bf x} \to {\bf x} + {\bf n} L$, where $L$ is
the size of the torus and ${\bf n}$ is a vector with integer
components. The corresponding Dirichlet domain is a cube (more
generally, a parallelepiped) with opposite faces identified (glued
together). This is, in fact, the model one is studying when one
simulates the universe in a finite box with periodic boundary
conditions.

Since the eigenfunctions of the Laplacian on $T^3$ are simply the
discrete plane waves, the evaluation of $\xic$ on $T^3$ as a sum over
modes functions (see eq.(\ref{moi1})) using Fast Fourier Transforms
is a preferred technique, one we have used extensively to constrain
the size of such models using the \cobedmr data.  However, we revisit this
simple case to illustrate the various steps and clarify subtleties
involved in the calculation of $\xic$ in a multiply connected universe
using the method of images.  The toroidal case also provides a good
benchmark for evaluating the efficiency of the method of images.

The correlation function in the periodic box implied by the $T^3$
topology is
\begin{equation}
\xi_\Phi^{T^3}({\bf x,x^\prime}) = \frac{1}{L^3} \sum _{{\bf n}}
P_\Phi(k_{\bf n}) \exp\left[-i \frac{2 \pi {\bf n}}{L}\cdot ({\bf x} - 
{\bf x}^\prime)\right]
\label{xic_tor}
\end{equation}
where ${\bf n} \equiv (n_x,n_y,n_z)$ is 3-tuple of integers,
$k^2_{\bf n} = (2\pi/L)^2 ({\bf n} \cdot {\bf n})$, and the term with
${\bf n}\cdot{\bf n}=0$ is excluded from the summation. This is a
direct consequence of substituting the known eigenmode functions of
the Laplacian on $T^3$, $\Psi_{\bf n}({\bf x}) = \exp(i 2\pi {\bf n} \cdot
{\bf x}/L)$, into eq. (\ref{xi1}) for the correlation function.

The method of images leads to the following alternative derivation of
(\ref{xic_tor}).  The correlation function on ${\cal E}^3$ is given by
\begin{equation}
\xiu = \frac{1}{(2\pi)^3} \int d^3k~P_\Phi(k) e^{i {\bf k} \cdot
({\bf x} - {\bf x} \prime)}\, . \\
\end{equation}
Assume, without loss of generality,\footnote{For any two points, ${\bf
x}$ and ${\bf x}^\prime$ in ${\cal M}^u$, there exists a discrete
motion $\gamma \in \Gamma$ such that the image $\gamma{\bf x}^\prime$,
would be in the Dirichlet domain with ${\bf x}$ as basepoint. Since
${\bf x}^\prime$ and $\gamma {\bf x}^\prime$ are equivalent for the
compact space, considering the compact space correlation between ${\bf
x}$ and $\gamma {\bf x}^\prime$ instead of ${\bf x}$ and ${\bf
x}^\prime$ will give identical results.}  that ${\bf x^\prime}$ lies in
the Dirichlet domain with ${\bf x}$ as the basepoint. The contribution
of each image has the form
\begin{equation}
\xiugj =\frac{1}{(2\pi)^3} \int d^3k P_\Phi(k) e^{i {\bf k} \cdot 
({\bf x} - {\bf x} \prime)} ~\exp\left[ i (k_x n^j_x + k_y n^j_y + k_z
n^j_z)\right]\,.
\end{equation}
from images $\gamma_i {\bf x}^\prime = {\bf x}^\prime + {\bf n}^j
L$. Summing over the contribution from the images, the correlation
function on $T^3$ is
\begin{mathletters}
\begin{eqnarray} \xi_\Phi^{T^3}({\bf x,x^\prime}) &=&\lim _{N\to\infty}
\frac{1}{(2\pi)^3} \int d^3k P_\Phi(k) e^{i {\bf k}\cdot ({\bf x}
-{\bf x} \prime)} \sum _{n_x=-N}^N \sum _{n_y=-N}^N\sum _{n_z=-N}^N
e^{i {\bf k} {\bf n} L} \nonumber\\ &=& \frac{1}{(2\pi)^3} \int d^3k
P_\Phi(k) e^{i {\bf k} ({\bf x} - {\bf x} \prime)}\lim _{N\to\infty}
\left[\frac{\sin\left[(N+1/2)k_x L\right]}{\sin(k_x
L/2)}\frac{\sin\left[(N+1/2)k_y L\right]} {\sin(k_y
L/2)}\frac{\sin\left[(N+1/2)k_z L\right]}{\sin(k_z
L/2)}\right]\lbl{tor_delapprox}\\ &=& \int d^3k P_\Phi(k) e^{i {\bf k}
({\bf x} - {\bf x} \prime)} \sum _{j=0}^{\infty} \delta ({\bf k}L - 2
\pi {\bf n}_j) \nonumber\\ &=&\frac{1}{L^3}\sum _{{\bf n}}
P_\Phi(\frac{2\pi}{L}{\bf n}) \exp\left[-i \frac{2 \pi {\bf n}}{L}
({\bf x} - {\bf x} \prime)\right]\,.
\label{xic_tor_moi}
\end{eqnarray}
\end{mathletters}
The final eq.(\ref{xic_tor_moi}) is the same as (\ref{xic_tor})
except that it contains a term $P({\bf n}=0)$ which is infinite for a
wide class of power spectra; \eg $P(k) \propto k^\alpha$, $\alpha <
0$, including those $\alpha$ for which the integral $\int_0^\infty dk
k^2 P(k)$ is convergent at $k=0$, \ie $\alpha > -3$.

The regularizing term can be easily calculated as well. Subtraction of
the volume integral over $N$ domains as in eq.(\ref{eq:series}) is
described by the following  substitution in eq.~(\ref{xic_tor_moi})
\begin{equation}
\lim_{N \to \infty} 
\sum _{n_x=-N}^N \sum _{n_y=-N}^N\sum_{n_z=-N}^N e^{i {\bf k} {\bf n}
L} \rightarrow \lim_{N \to \infty} \left[1-j_0(k_x
L/2)j_0(k_yL/2)j_0(k_z L/2)\right] \sum _{n_x=-N}^N \sum
_{n_y=-N}^N\sum _{n_z=-N}^N e^{i {\bf k} {\bf n} L}
\label{xic_tor_moi_regsub}
\end{equation}
(where $j_0(x)$ is the zeroth order spherical Bessel function) which
leads to the following form for the regularized correlation function
\begin{equation}
\tilde\xi_\Phi^{T^3}({\bf x,x^\prime}) = \frac{1}{(2\pi)^3}\int d^3k
P_\Phi(k) e^{i {\bf k} ({\bf x} - {\bf x} \prime)} \left[ \sum _{|{\bf
n}_j| \ne 0}^{\infty} \delta ({\bf k}L - 2 \pi {\bf n}_j) +
\frac{(kL)^2}{24}\delta ({\bf k}L)\right]\,.
\label{xic_tor_moi_reg}
\end{equation}
As long as the power spectrum $P_\Phi(k)$ does not blow up faster that
$k^{-2}$ as $k \to 0$, the above regularization removes the zero mode
contribution completely. In the case of the Harrison-Zeldovich
spectrum (equal power per logarithm of $k$), where formally $P_\Phi(k)
\sim k^{-3}$ as $k \to 0$, the regularization suppresses the $k=0$
contribution but does not eliminate it completely. However, any
physically motivated origin of $P_\Phi(k)$ such as from inflation does
have an infrared cut-off, and all that is required of the
regularization scheme is suppression of $k=0$ power.

Ability to perform an analytic summation over all the images in $T^3$
would lead to the exact recovery of the positions of the discrete
eigenvalues, $2\pi|{\bf n}|/L$, and the eigenfunctions in this compact
space.  In more complex topologies (\eg the CH spaces) one can only
sum over a finite number of images and estimate the limit from that.
If one tiles the universal cover out to $N$ layers, one recovers the
delta functions at ${\bf k} = 2\pi{\bf n}/L$ only approximately in the
partial sum over images. The power of each discrete mode is aliased to
a cubic cell of the reciprocal lattice\footnote{
$\int_{2\pi(m-\half )/L}^{2\pi(m+\half)/L} dk \sin[(N+1/2)k L]/\sin[kL/2] =
2\pi/L$ for any integer $m$ and $N$.}. In each cell, the kernel within
square brackets in eq. (\ref{tor_delapprox}) has a peak at $k=
2\pi\vert {\bf n} \vert /L$ of height $(2N+1)^3$ and width $\approx
1/N$, with damped oscillatory wings. Accurate regularization of the
partial sum then consists of subtracting the total power in the
reciprocal lattice cell around ${\bf k}=0$.

The left panels of Figure ~\ref{fig_tor_moi} show the number density
of states $n^{T^3}(k)/k^3$ obtained by summation over $N=9$ layers of
images.  The topmost panel shows the direct sum while the middle panel
illustrates the effect of the regularizing term. The regularization
procedure drastically reduces the spurious power below the fundamental
frequency. The oscillatory wings of power aliasing inside each
reciprocal cell can be eliminated if one averages over the results at
each $k$ obtained at different values of $N$. We found it is most
effective to do it by Cesaro resummation: the effect is shown in the
bottom panel.  The result is a smoothed approximation to the
underlying discrete spectrum (marked by arrows in the figure) with
negligible contribution below the fundamental frequency and a
positively defined spectrum.

The plots of the right panel of Figure~\ref{fig_tor_moi} are analogous
to those of the left panel except that they use the radial limiting
procedure described in eq.~(\ref{moi_final}) below, with $r_* =
9L$. The radial limiting procedure does better at eliminating the
unwanted low $k$ power at the resummation stage. The spectral lines
are somewhat broader solely due to the fact that there are about
$\pi/6$ times fewer images within a sphere of radius $r_*
=9L$ than in the $N=9$ layers.

As we mentioned above, the regularization is not strictly complete for
Harrison-Zeldovich like spectra, so the question of the numerically
superior technique does arise. The resummation procedure effectively
averages the sequence of partial sums up to a given distance.  The
superior low $k$ power elimination of the radial limiting procedure is
related to the fact that the product of the volume $L^3$ and the
number of images within a radius $r$ jitters around the volume $4\pi
r^3 /3$ on very short scales. The unwanted residual power in low $k$
modes is distributed in this jitter, and is more readily removed by
Cesaro resummation even if one does not go far in $r$. In the other
case, the residual power in low $k$ modes is distributed in a more
orderly wave of wavelength $\sim k^{-1}$ in the sequence of partial
sums. Hence, when one is summing images up to a finite distance $d$ 
the residual power in modes with $k d \ll 1$ is not averaged out.

\begin{figure}[h]
\plottwoside{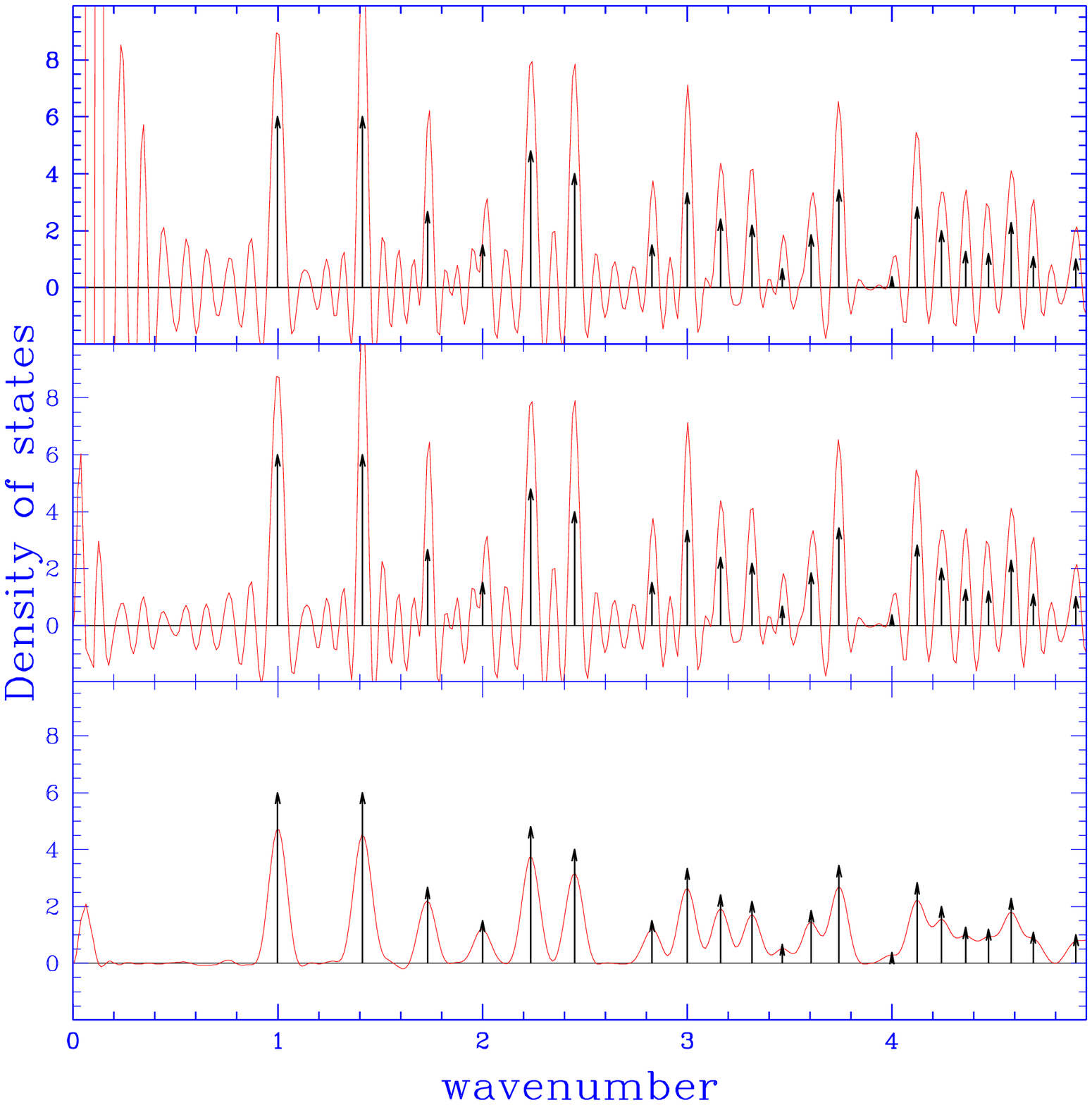}{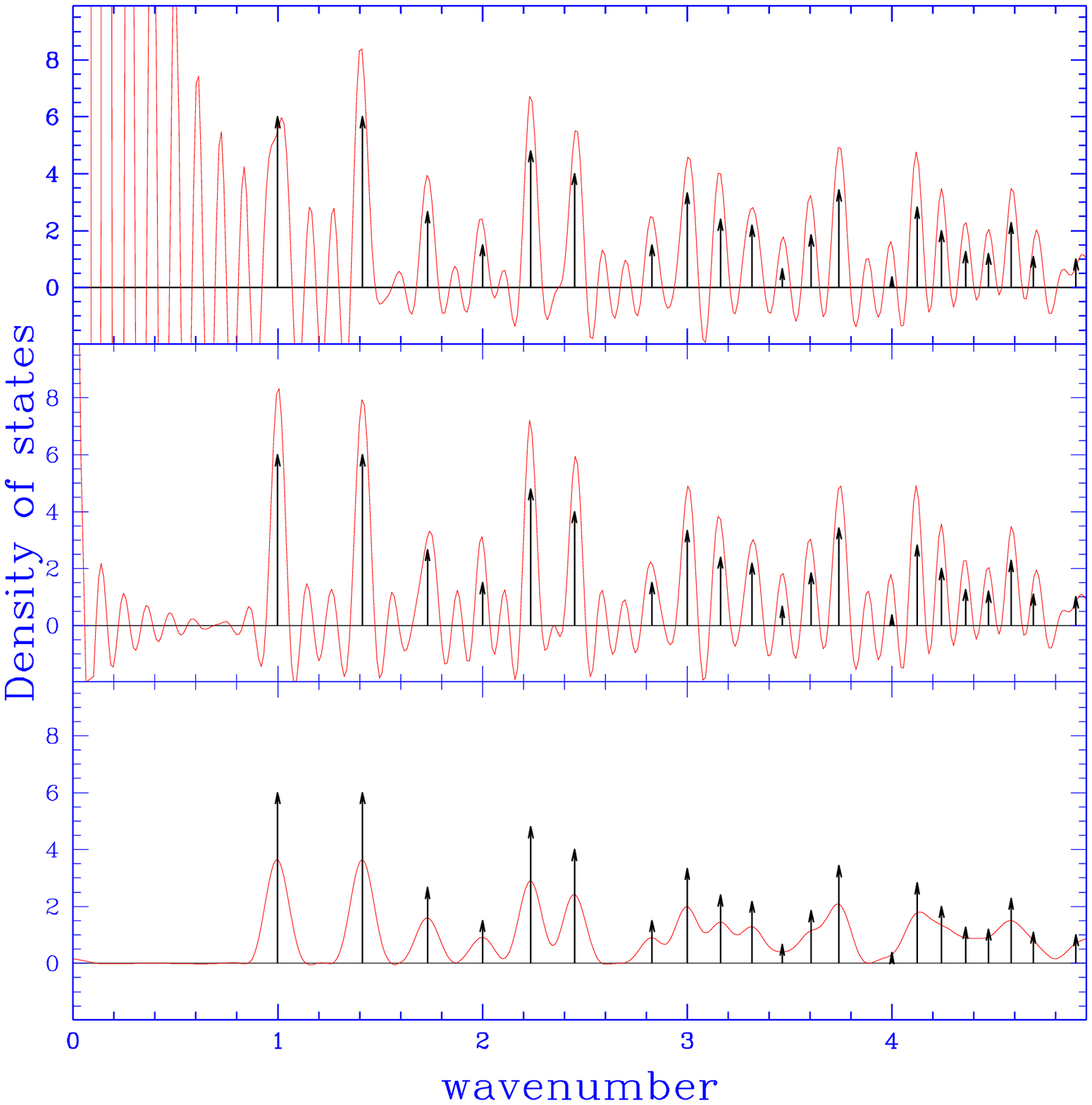}
\caption{This illustrates the recovery of the known discrete spectrum
in the $T^3$ model of size $L=2\pi$ using the method of images. Plots
of $(2\pi^2/k^3)~n^{T^3}(k)$ in the three panels show the density of states
obtained at different stages of our method. The {\em left panels}
implement the limiting procedure of eq.(\ref{eq:series}), varying the
number of shells $N$ of images that are summed. Images up to $N=9$
layers ($=19^3$ images ) are used. The {\em right panels} use the
radial limiting procedure of eq.(\ref{moi_final}), with a cut-off
radius $r_*$ that we use for CH spaces; here, $r_* =9L$, with only
$3071$ nearest images used.  The vertical arrows mark the location of
the discrete eigenvalues with the height proportional to the
occupation number. The {\em topmost panel} shows the unregularized
spectrum. Note the huge spurious contribution from the $k=0$ mode
below the fundamental mode at $k=1$. The {\em middle panel} shows the
result of regularization. At this stage most of the spurious power is
removed. The result of Cesaro resummation shown in the {\em bottom
panel} demonstrates that our method finally recovers a smoothed power
spectrum with negligible spurious long wavelength power. Note that the
radial limiting procedure does better in terms of removing spurious
power at low $k$.}
\label{fig_tor_moi} 
\end{figure}

It is important to realize that at a given partial image sum the
correlation function is obtained to far better accuracy than the power
spectrum. The correlation function is a $k$-space integral over the
power spectrum and is accurately reproduced as long as power is
sufficiently peaked around the correct eigenvalues and there is not
much overlap between the adjacent peaks. The partial image sum
generates a spectrum which can be approximated as the true spectrum
convolved with a smearing function, $W((k- k_i)^2/ (\Delta k)_s ^2)$,
around the true eigenvalues. Here $(\Delta k)_s$ quantifies the width
of the smoothing. In a compact space no two points are physically
separated by more than the diameter of the space. Hence, to get a
reasonably accurate estimation of the correlation function, it is
sufficient to ensure that $R_s \sim 1/(\Delta k)_s$ is larger than the
diameter of the space.  In the case of the simple $T^3$ model, in the
Gaussian approximation to the smearing function $W$, the method of
images with $N$ layers gives $R_s = (2/3)d_T\sqrt{N(N+1)} $ where $d_T
= L \sqrt{3}/2$ is the diameter of the simple $T^3$. Hence, the second
layer ($N=2$) approximation satisfies the $R_s > d_T$ condition by a
comfortable margin, and even the first layer comes fairly close.

The above estimate of the spread in the peaks around eigenvalues is
based solely on the regularized spectrum done with summation in layers 
up to $N$ layers. We also carry out a Cesaro resummation to remove
remaining spurious low $k$ power. The resummation procedure
effectively averages the cumulative results at each layer. It is not
difficult to see that this reduces the amplitude of the peak by a
factor of two and broadens the spectrum by the same
factor. Thus to reach the same spectral resolution after Cesaro resummation
one need to go to $N_c \approx 2N$ layers.
We prefer to implement the radial limiting
procedure, which is simpler and does better at the resummation stage in
suppressing residual low $k$ power. Equating the number of images of
the radial limiting procedure to that in $N_c$ layers implies choosing
$r_*/L \approx (3/4\pi)^{1\over3} (2 N_c+1)$. Hence, for the resummed
spectrum obtained with the radial limiting procedure, the comfortable
margin of $N=2$ (which translates to $N_c=4$)
estimated above for obtaining accurate correlation
functions translates to $r_* \approx 5.6 L$. In the second panel of
Figure~\ref{fig_tor_moi2}, we show the spectrum recovered at $r_*
= 5.5 L$.
\begin{figure}[h]
\plotone{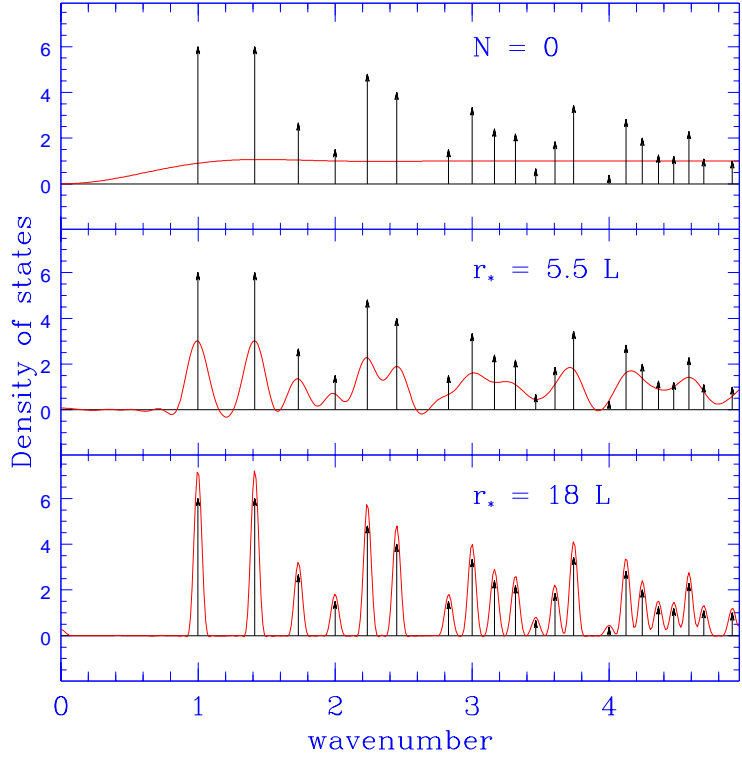}
\caption{The recovery of the known discrete spectrum 
$(2\pi^2/k^3)~n^{T^3}(k)$ in the $T^3$
model of size $L=2\pi$ using the method of images. The top panel shows
the regularized spectrum obtained at the lowest level approximation
($N=0$, nearest image): only the low $k$ cutoff is recovered. The
second panel is the spectrum recovered after summing images up to a
distance of $r_* = 5.5L$, a level of approximation which should (and
does) recover the correlation function quite well, although the
spectrum is coarse. The bottom panel demonstrates the convergence of
the regularized method of images: with a large enough number of
images, the spectrum is also recovered with precision. In the
lower two panels, Cesaro resummation has been carried out on the
regularized spectrum. }
\label{fig_tor_moi2} 
\end{figure}

Another important remark is in order. In application to CMB, the
positions of points ${\bf x},{\bf x}^\prime$, between which the
correlation function is to be calculated, are given by the length and
directions of the photon path from the observer, \ie correspond to the
coordinates on the universal cover ${\cal M}^u$.  In the method of
images, the value of $\xic$ for the pairs of points belonging to
separate domains \footnote{More precisely, when ${\bf x}^\prime$ does
not belong to the Dirichlet domain constructed around ${\bf x}$.}  is
found by symmetric replication of the correlation value computed after
${\bf x}^\prime$ is mapped back into the Dirichlet domain around ${\bf
x}$.  Thus, the method of images applied to the correlation function
preserves at all levels of approximation the exact periodicity of the
$\xic$ viewed as being defined on the universal cover.  This periodicity
is the major distinction between $\xic$ and the correlation function
$\xiu$ in the simply connected universal cover space.  In contrast,
a correlation function determined as the inverse transform with
respect to the universal cover eigenmodes of the approximate power
spectrum will fail to obey the symmetries of the compact tiling
strictly. This failure is greater the cruder the level of
approximation used for computing the power spectrum.

The success at reconstructing the correlation function by the method
of images with even just a few layers of summation is demonstrated in
Figure~\ref{fig_xiT3}.  In accordance with the estimate of the
convergence of the method, summation over 5 layers (N=5.5) and above
produces a correlation function almost indistinguishable from the
exact result.  $N=3.5$ is already very good and even $N=1.5$ is an
adequate approximation of the result.  Even from the very beginning,
with only the nearest image used, the method of images correctly
catches the qualitative behavior of the correlation function in the
compact space, which is dramatically different from the corresponding
correlation function in non-compact flat space.
\begin{figure}[h]
\plotone{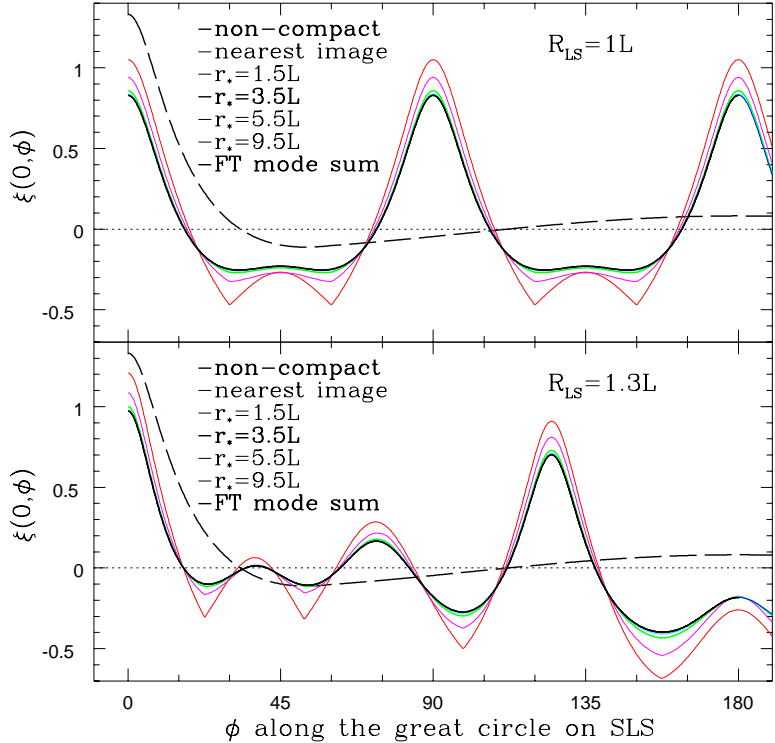}
\caption{Correlation function in $T^3$, calculated by the method of
images with successively increasing number of images, is compared with
the exact summation of eigenmodes of the Laplacian via Fourier
transformation.  Anticipating application to the CMB, we plot the
correlation values between the points along the great circle of radius
$R_{\sc ls}$ and one fixed point on the circle at $(R_{\sc ls},0,0)$.
The orientation of Cartesian coordinates coincides with directions of
periodicity of the torus, so $\phi$ is the polar angle. Monopole and
dipole contributions {\it along the circles} (\ie different on each
circle) are subtracted.  The results are shown for two values of
$R_{\sc ls}$: The first case when $R_{\sc ls}=1$ is a very symmetric
one where points at $0,90,180,\ldots$ degrees on the great circle are
exact images of the point at the origin.  The second case is a more
general one. Labels mark the curves from top to bottom at zero lag,
$\phi=0$.}
\label{fig_xiT3} 
\end{figure}

\section{Compact Hyperbolic spaces}
\label{sec_CH}

\subsection{Correlation function on $\hm$}
\label{ssec_hypcorr}

The local isotropy and homogeneity of $\hm$ implies $\xiu$ depends
only on the proper distance, $r\equiv d({\bf x},{\bf x^\prime})$,
between the points ${\bf x}$ and ${\bf x^\prime}$.  The eigenfunctions
on the universal cover are of course well known for all homogeneous
and isotropic models~\cite{har67}. Consequently $\xiu$ can be obtained
through eq. (\ref{xi1}).  The role of $\xiur$ in the compact
space calculation is to impose the desired power spectrum of $\Phi$.
In line with the application for which we developed our method, we now
specialize to the scalar field $\Phi$ being the one describing
cosmological gravitational potential fluctuations.

The initial power spectrum of the gravitational potential $P_\Phi(k)$
is believed to be dictated by an early universe scenario for the
generation of primordial perturbations. We assume that the initial
perturbations are generated by quantum vacuum fluctuations during
inflation. This leads to
\begin{equation} 
\xiu \equiv \xiur = \int_0^\infty
\frac{d\beta~\beta}{(\beta^2 +1)} ~\frac{\sin(\beta r)}{\beta \sinh
r}~~{\cal P}_\Phi(\beta) \, , \lbl{xiur}
\end{equation}
where $\beta\equiv\sqrt{(k d_c)^2 -1}$, ${\cal P}_\Phi(\beta)\equiv
\beta(\beta^2+1) P_\Phi(k)/(2\pi^2)$ and, as before, $r$ is in units
of $d_c$. In the simplest inflation models, the power per logarithmic interval of
$k$, \ie ${\cal P}_\Phi$, is approximately constant in the \bqt
subcurvature sector\eqt, defined by $k d_c > 1$. This is the
generalization of the Harrison-Zeldovich spectrum in spatially flat
models to hyperbolic spaces~\cite{lyt_stew90,rat_peeb94}.  In
appendix~\ref{app_genpert}, we outline a simplified calculation of the
initial inflationary perturbation spectra ${\cal P}_\Phi$ in
hyperbolic geometry and derive a broader class of ``tilted'' spectra.
Sub-horizon vacuum fluctuations during inflation are not expected to
generate supercurvature modes, those with $k d_c <1$, which is why
they are not included in eq.(\ref{xiur}). Indeed, since $H^2 > 1/(a
d_c)^2$, for modes with $k d_c < 1$ we always have $k/(aH) < 1$ so
inflation by itself does not provide a causal mechanism for their
excitation. Moreover, the lowest non-zero eigenvalue, $k_1>0$ in
compact spaces provides an infra-red cutoff in the spectrum which can
be large enough in many CH spaces to exclude the supercurvature sector
entirely ($k_1 d_c > 1$).  (See $\S$\ref{ssec_CH_powspec}.)  Even if
the space does support supercurvature modes, some physical mechanism
needs to be invoked to excite them, \eg as a byproduct of the creation
of the compact space itself, but which could be accompanied by complex
non-perturbative structure as well.  To have quantitative predictions
for $P_\Phi(k)$ would require addressing this possibility in a full
quantum cosmological context. For a recent discussion of the creation
of hyperbolic universes within a quantum cosmological framework
see~\cite{CHcreat}.

\subsection{Numerical implementation of the method of images}
\lbl{ssec_CH_numer}

Equation (\ref{moi1}) encodes the basic formula for calculating the
correlation function using the method of images. For a numerical
estimate, a limiting procedure such as eq.~(\ref{eq:series}) has to be
used. In this form, the regularization term involves integrating $\xiu$
over a Dirichlet domain. Such terms can be numerically computed given
the discrete group of motion, $\Gamma$, but it is usually cumbersome
since the Dirichlet domains of CH spaces have complicated shapes which
vary depending on the basepoint.

A more effective and simpler radial limiting procedure in implementing
the regularized method of images is to explicitly sum images up to a
radius $r_*$ and regularize by subtracting the integral $\xiur$ over a
spherical ball of finite radius $r_*$:
\begin{equation}
\xic = \lim_{r_*\to\infty} \left[ \sum_{r_j < r_*} \xiurj - 
\frac{4\pi}{V_{\!\cal M}} \int_0^{r_*} dr ~\sinh^2r ~\xiur\right],~~~~
r_j = d({\bf x}, \gamma_j {\bf x}^\prime) \le r_{j+1}\, . 
\lbl{moi_final}
\end{equation}
The volume element in the integral is the one for $\hm$. We have shown
for the flat torus that this scheme works better numerically than 
formal regularization which subtracts integrals over Dirichlet domains.

The plot on the left in Figure~\ref{fig_moi} illustrates the steps
involved in implementing the regularized method of images.  The value
of $\xi^c_\Phi$ as a function of $r_*$ has some residual jitter, which
arises because of the boundary effects due to the sharp \bqt top-hat\eqt
averaging over a spherical ball chosen for the counterterm in
eq.(\ref{moi_final}).  This can be smoothed out by resummation
techniques~\cite{hardy}.  We use Cesaro resummation for this purpose.

In hyperbolic spaces, the number of images within a radius $r_*$ grows
exponentially for $r_*/d_c > 1$ and it is not numerically feasible to
extend direct summation to large values of $r_*/d_c$. The presence of
the counterterm, however, besides regularizing, significantly improves
convergence. This can be intuitively understood as follows: $\xiurj$
represents a sampling of a smooth function at discrete points
$r_j$. In a distant radial interval $[r,r+dr],~r \gg R_>, ~dr \sim
R_>$, there are approximately $ (4\pi/V_{\!\cal M})~\sinh^2r~dr$
images.  The sum, $\sum_{r_j} \xiurj$, within this interval is similar
to the (Monte-Carlo type) estimation of the integral, therefore one
may approximate the sum over all distant images beyond a radius $r_*$
by an integral to obtain
\begin{equation}
\tilde\xic =  \sum_{r_j < r_*} \xiurj + 
\frac{4\pi}{V_{\!\cal M}} \int_{r_*}^\infty dr~ \sinh^2r ~\xiur\,.
\lbl{moi3}
\end{equation}
The tilde on $\xic$ denotes the fact that it is approximate and
unregularized. Subtracting the integral $(4\pi/V_{\!\cal
M})\int_0^\infty dr ~\sinh^2r ~\xiur $ as dictated by the
regularization eq.(\ref{moi1}), we recover the finite $r_*$ term of
the limiting sequence in eq.(\ref{moi_final}).  This demonstrates that 
even at a finite $r_*$, in addition to the explicit sum over images
with $r_j<r_*$, the expression for $\xi^c_\Phi$ in
eq.(\ref{moi_final}) contains the gross contribution from all distant
images with $r_j > r_*$. Numerically we have found it suffices to evaluate
the above expression up to $r_*$ about $4$ to $5$ times the domain
size $R_>$ to obtain a convergent result for $\xic$.

\begin{figure}[htb]
\plottwoside{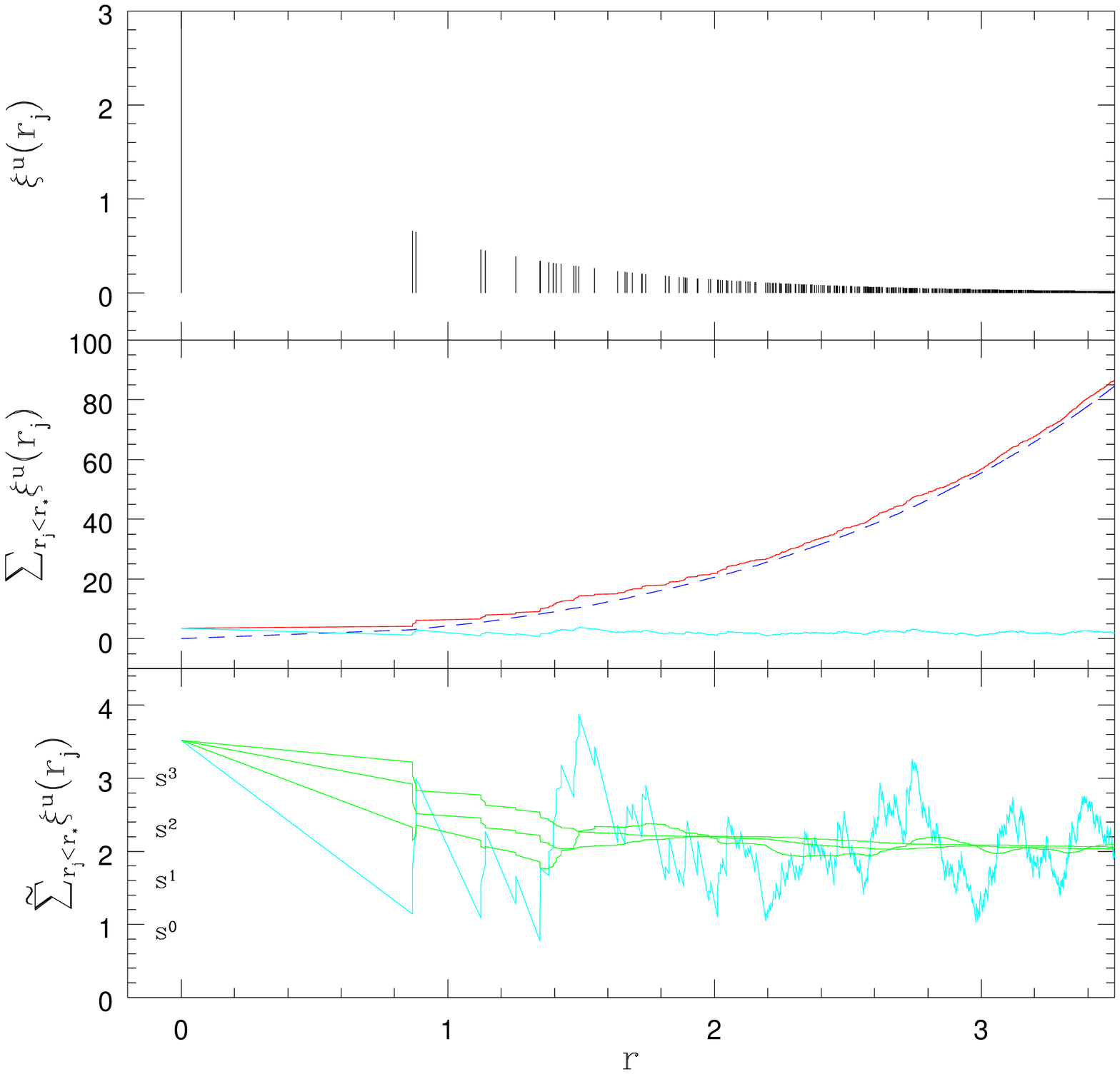}{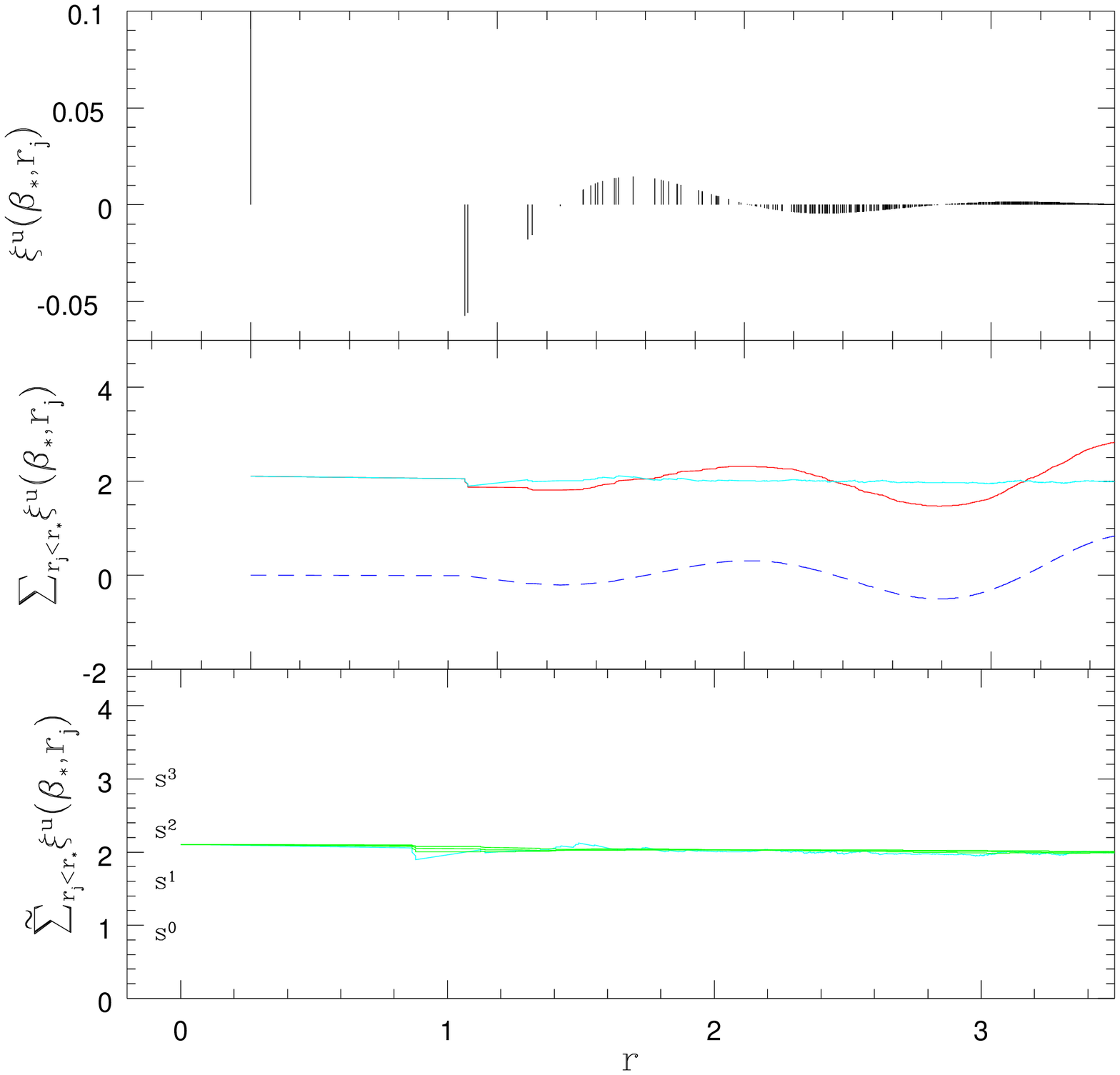}
\caption{The two plots illustrate the regularization of the
correlation functions with and without an infrared cut-off at $\beta_*
> 0$. The example shown is the correlation function at zero-point
separation at some point on the CH manifold m004(-5,1).  The plot on
the left corresponds to $\xic$ where there is no $\beta$ cut-off.  The
topmost panel shows the sampled values of $\xiur$ which contribute to
the sum over images.  The upper solid curve in the middle panel shows
the divergent cumulative build up of the partial sum over images with
successive addition of distant images. The dashed curve is the
regularizing counterterm required to remove the zero-mode contribution
and the lower solid curve is the cumulative value of the regularized
partial sum $\tilde\xic$, which fluctuates around the true value once
a sufficient number of images have been added. In the bottom panel
this residual jitter ($s^0$) in the estimate of the correlation is
removed by Cesaro resummation. The sequence of lines, $s^1,s^2$ and
$s^3$ shows the result of applying first, second and third order
Cesaro resummation. The accuracy at the second order is usually
sufficient.  The plot on the right is analogous to the left but for
the auxiliary correlation function $\xicaux$ with an infrared cut-off
at $\beta_*=4.0$ below the first eigenvalue in the CH space. In
contrast to the left panels, the cumulative sum over images is
oscillatory which is more easily regularized, leading to much smaller
residual jitter around the true value.  }
\label{fig_moi} 
\end{figure}

Eq.~(\ref{moi3}) provides the simplest interpretation of our
regularization procedure as an integral approximation to the total
contribution of all distant images outside the region over which
direct summation has been carried out.  However, this interpretation
may not be obvious in all cases. In fact, for the spectrum ${\cal
P}(\beta)\equiv\mathrm{const}$ that we use here, the correlation
function on the universal cover $\hm$ is given in terms of the
hyperbolic sine and cosine integral functions, $\shint(r)$ and
$\chint(r)$, respectively, by
\begin{equation}
\xiur = \chint(r) - \coth(r)\,\shint(r) \, . 
\lbl{xiur_anal}
\end{equation}
This is positive definite and does not fall off fast enough with $r$ for its
volume integral to converge. As a result, the integral in
eq.(\ref{moi3}) is not defined and the step from
eq.(\ref{moi_final}) to eq.(\ref{moi3}) is nontrivial, involving the
regularization of an infinity which can be traced to the $\beta =0$
mode. To reinstate the intuitive interpretation in this case, we first
show that it is valid for an auxiliary correlation function, $\xiaux$,
which has an explicit infrared cutoff at $\beta_* > 0$. The auxiliary
correlation function
\begin{eqnarray}
\xiaux &=& \int_{\beta_*}^\infty
\frac{d\beta~\beta}{(\beta^2 +1)} ~\frac{\sin(\beta r)}{\beta \sinh
r}~~{\cal P}(\beta) \nonumber \\ 
&=& -{\rm Re}\left\{\left[ \cint\left((i-\beta_*)r\right) +
\cint\left((i+\beta_*)r\right)\right]/2 + i \left[
\sint\left((i+\beta_*)r\right) + \sint\left((i-\beta_*)r\right)
\right]/2\coth r \right\}, 
\lbl{xiur_aux}
\end{eqnarray}
where $\sint(r)$ and $\cint(r)$ are the sine and cosine integral
functions, respectively.  This function, $\xiaux$, is no longer
positive definite and its volume integral, although an improper
integral, can be shown to be zero. 

The contribution to $\xicaux$ of the distant images with $R\gg\beta_*^{-1}$ 
can now be evaluated explicitly:
\begin{equation}
4\pi\int_R^\infty dr ~\sinh^2r~\xiaux \stackrel{ R
\gg\beta_*^{-1}}{\longrightarrow} -\frac{4\pi}{V_D k_*^3}
\sin\left(\beta_*R + \sin^{-1}(k_*^{-1})\right)~\frac{\sinh R}{R}\, . 
\lbl{intxiaux}
\end{equation}
This is purely oscillatory with a zero mean. The plot on the right in
Figure~\ref{fig_moi} shows that these oscillations of the
regularization term precisely cancel out the oscillations (with
growing amplitude) of the image sum, with no net effect on the
limiting value of $\xicaux$.

Having established that eq.~(\ref{moi3}) makes sense for $\xicaux$, 
where $\beta_* > 0$, the interpretation may now easily be extended to
$\xic$ by simply noting that given $0< \beta_* \ll \beta_1$, where
$\beta_1$ is the wavenumber corresponding to the first eigenvalue of
the Laplacian in the CH space, the method of images applied to the
auxiliary $\xiaux$ and $\xiu$ must converge to the same value.  This
result is demonstrated in Figure~\ref{fig_app_moi}.

\begin{figure}[h]
\plotone{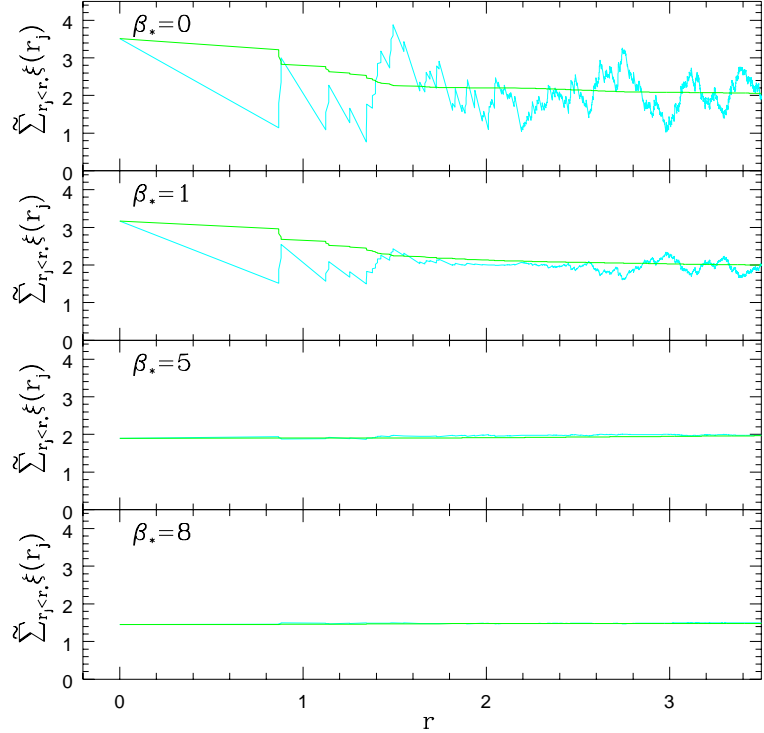}
\caption{The analog of the bottom panels of
Figure~\ref{fig_moi} for three values of the cut-off $\beta_*$.
The more jagged curve is a regularized sequence of partial sums and the
smoother curve is the result of a third order Cesaro resummation on
the regularized result. The effectiveness of regularization is
enhanced (reflected in the smaller amplitude of jitter) as one
increases the value of $\beta_*$ by directly cutting out more of the
zero-mode contamination. The computation is robust and converges to
the correct value as long as $\beta_* \ll \beta_1$.  For the CH
manifold m004(-5,1) used in this figure, we estimate $\beta_1 \gta 5$ (see
Figure~\ref{fig:kcutoffs}). In the top three cases, $\beta_*\le 5$,
and the computed correlation converges to $\xi^c(0) =2$ with high
accuracy,  independent of $\beta_*$. When $\beta_* > \beta_1$, the value
of $\xi^c(0)$ is underestimated, as is seen in the bottom most panel
where $\beta_*=8$.}
\label{fig_app_moi} 
\end{figure}

\subsection{Power spectrum }
\lbl{ssec_CH_powspec}

In this section, we present our results on the power spectrum of CH
manifolds obtained by applying the method of images to the density of
states kernel $n^u(k,{\bf x}, {\bf x}^\prime)$ discussed in
$\S$\ref{ssec_powspec}. We emphasize that evaluating the density of
states numerically in CH spaces is not our primary goal. What we
present is a rough estimate of the power spectra in CH spaces that can
be readily obtained as a `byproduct' of our primary goal of computing
correlation functions in CH spaces.

Indeed, in the formalism of images, the singular delta-functions in
$n^c(k)$ (see eq~(\ref{cct_powspec})) are recovered, in principle, by
the precise cancellation of the smooth contributions from {\em all}
images. The volume of a sphere in $\hm$, and consequently the number
of images (of the Dirichlet domain of a CH space on the universal
cover), grows exponentially with radius beyond $r \sim d_c$. This is
the primary constraint on the success of applying the method of images
to the density of states.

Our approximation for computing the CH space correlation function
includes the gross integral estimate of the impact of distant images
which results in a spread-out convolved density of states
distribution. Increasing $r_*$ progressively sharpens spectral
profiles near the true positions of the discrete
eigenvalues~\cite{bal_vor86}.  Figure~\ref{fig:kcutoffs} shows a
sample $n^c(k,{\bf x},{\bf x})$ for some of the CH manifolds at two
random positions ${\bf x}$ in the first two columns. The third column
shows the smoothed out estimation of the actual density of states,
$n^c(k)$, for the CH manifolds that one can obtain by averaging
$n^c(k,{\bf x},{\bf x})$ over $\sim 10^3$ points on the manifold.

There is a definite signature of strong suppression of power at small
$\beta$ in all the cases that we have explored.  This is qualitatively
similar to the infrared cutoff known for the compact manifolds with
flat and spherical topology.  Quantitatively, the break appears around
$k \sim {\cal O}\left(d_{\!\cal M}^{-1}\right)$, consistent with the
intuitive expectations.
\begin{figure}[htb]
\plotone{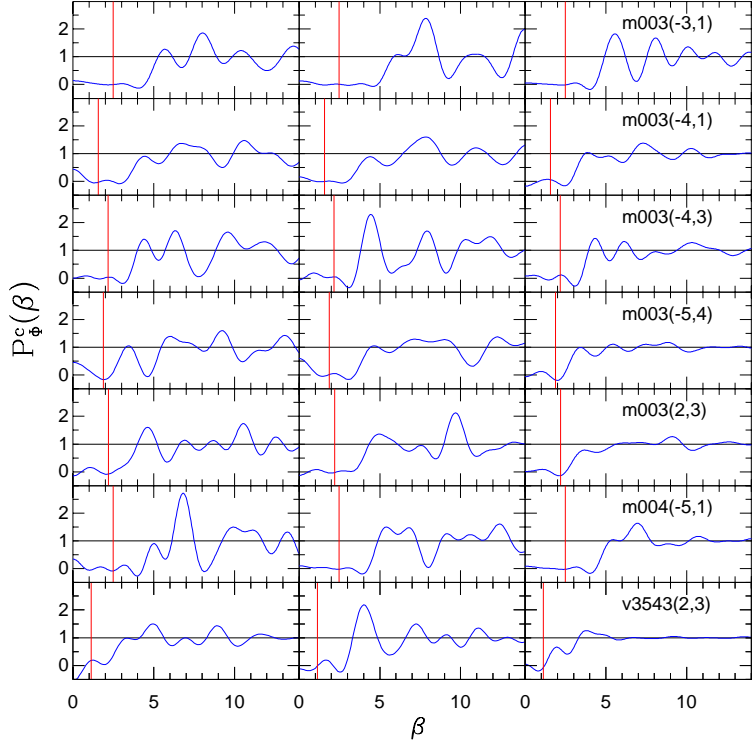}
\caption{ The density of states exhibits suppression of long-wave
power if the Universe has compact topology.  Each row corresponds to
one compact hyperbolic space, indicated by the label in the last
column. Left and middle columns show $n^c(k,{\bf x,x})/k^3$ computed
with the method of images at two randomly selected points ${\bf
x}$. The last column shows the spatial averaged $n^c(k)/k^3$
obtained by Monte-Carlo integration of $n^c(k,{\bf x,x})/k^3$ over the
Dirichlet domain. The normalization is chosen such that the functions
are equal to unity for the topologically trivial infinite open
Universe.  Vertical lines illustrate naive estimation of the cutoff at
$k\sim \pi/d_{\cal M}$ which holds surprisingly well for the cases
checked.  The first 6 examples of the CH space have about the same
value of $d_{\cal M}$. These spaces have volumes $V/d_c^3$ around
unity. The sixth space is m004(-5,1), the \bqt small\eqt space used in
many of our examples. The seventh space is v3543(2,3), with volume
$V_{\!\cal M}/d_c^3= 6.45$, used as an example of a \bqt large\eqt
space in this paper. }
\label{fig:kcutoffs}
\end{figure}

An infra-red cutoff at the lowest non-zero eigenvalue, $k_1 >0$,
exists for all compact spaces.  {\em Cheeger's
inequality}~\cite{cheeg70} provides a lower bound on $k_1^2$ for a
compact Riemannian manifold, $M$:
\begin{equation}
k_1 \ge \frac{h_C}{2}, ~~~~h_C= \inf_{S}~\frac{A(S)}{min
\{V(M_1),V(M_2)\}} \, , 
\lbl{cheeg_ineq}
\end{equation}
where the infimum is taken over all possible surfaces, $S$, that
partition the space, $M$, into two subspaces, $M_1$ and $M_2$, i.e.,
$M=M_1\cup M_2$ and $S=\partial M_1 =\partial M_2$ ($S$ is the
boundary of $M_1$ and $M_2$).  {\em Cheeger's isoperimetric
constant} $h_C$ depends more on the geometry than the topology of the
space, with small values of $h_C$ achieved for spaces having a
\bqt dumbbell-like\eqt structure -- a thin bottleneck which allows a
partition of the space into two large volumes by a small-area
surface~\cite{cheeg70,bus80}. Regular shaped compact spaces do not
allow eigenvalues which are too small. For example, the Cheeger limit
for all flat $T^3$ manifolds is $k_1 \ge 2/L$, where $L$ is the
longest side of the torus.  Although the direct estimation of $h_C$ is
not simple, for any compact space ${\cal M}$ with curvature bounded
from below there exists a lower bound on $h_C$ in terms of the
diameter of ${\cal M}$~\cite{ber80,chav84}; for a 3-dimensional CH
space,
\begin{equation}
k_1 \ge h_C/2 \ge \frac{1}{d_{\!\cal M}}~ {\left[2\int_0^{1/2}\!\!dt
\cosh^2(t)\right]}^{-1}= 0.92/d_{\!\cal M}.  \lbl{cheegmin}
\end{equation}
This result prohibits supercurvature modes for all CH spaces with
$d_{\!\cal M} < 0.92\,d_c$.

There are other lower bounds on $k_1$ that exist in the literature.  In
terms of the diameter alone, the bound $k_1^2 \ge \pi^2/(2d_{\!\cal
M})^2-  {\mathrm max}\{-(n-1)K,0\}$ has been derived for any compact
$n$ dimensional space, with $K=0$ for flat geometry and $K = \pm
d_c^{-2}$ for spherical and hyperbolic geometries,
respectively~\cite{li_yau80}. For hyperbolic spaces, this bound is
sharper than the one above for $d_{\!\cal M} \lta 0.9\,d_c$. There is
another lower bound on $k_1$ in terms of the volume as well as the
diameter~\cite{yau75}; for CH spaces $k_1 \ge {\cal V}_{\!\cal M}/[2\pi
d_{\!\cal M}(\sinh(2 d_{\!\cal M}/d_c) -2 d_{\!\cal M}/d_c)]$. This
lower bound dominates in the supercurvature sector for volumes larger
than $\approx 7 d_c^3$.

\begin{figure}[htb]
\plotone{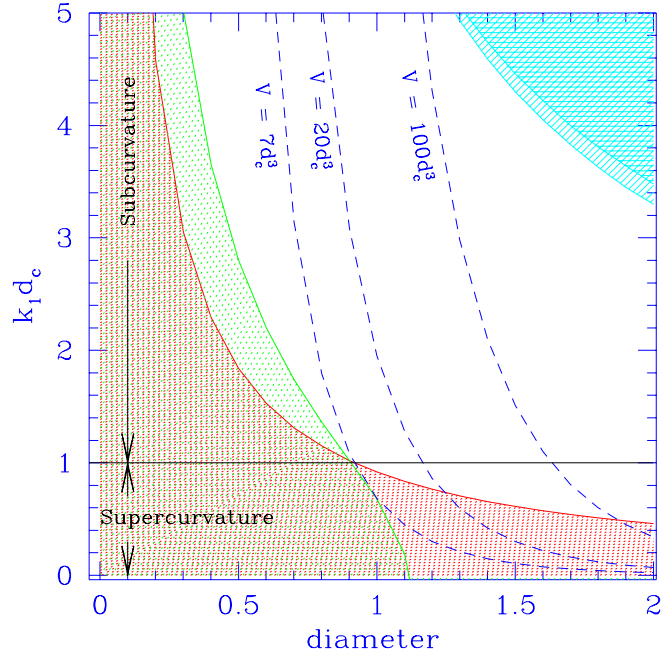}
\caption{Some known bounds on the first non-zero eigenvalue in CH
spaces are shown. The heavily shaded region is the lower bound in
eq.~(\ref{cheegmin}). The lightly shaded region is the lower bound
derived in [40]. The dashed lines show the lower bound
which depends on volume as well for three values of the volume. The
shaded regions in the top right corner define two upper bounds on
$k_1$.  The horizontally shaded region is the upper bound coming from
comparison with Dirichlet eigenvalues [42]. The other region
is the sharpest possible bound that can arise using the upper bound in
terms of $h_C$ [43] and the lower bound on $h_C$ in
eq.~(\ref{cheegmin}). }
\label{fig:k1min}
\end{figure}

There are also upper bounds on $k_1$. The bound, $k_1^2 \le 4 h_C/d_c
+ 10 h_C^2$~\cite{bus82}, does not allow for a firm conclusion that
the space supports supercurvature modes, $k_1 d_c < 1$, unless
$d_{\!\cal M} \ge 10.6\,d_c$ (using eq.~(\ref{cheegmin})).  Upper
bounds based on comparison with the first Dirichlet
eigenvalue~\cite{cheng75} on a subdomain of ${\cal M}$ cannot impose
$k_1 d_c <1$, since Dirichlet eigenvalues cannot be less than
$d_c^{-1}$.

The density of states $n^c(k)$ defines the eigenvalue spectrum of the
Laplacian on a compact space. It is well known that there exist very
strong connections between the geometry and the topology of the space
and the spectral properties of the Laplacian.\footnote{It is widely
believed that, in two and three dimensions, the inverse problem of
identifying a space from the spectrum of the Laplacian is {\em
spectrally rigid.}, \ie there can be only a finite number of spaces
which have the identical spectrum (modulo symmetries)~\cite{gur80}. In
higher dimensions, there are known exceptions, but spectral rigidity
is quite generic. } As discussed above, the infra-red cut off in the
spectrum is broadly determined by the size (diameter) of the space and
its volume.

The {\em Weyl formula} is an example of a general and powerful result
connecting number of states to the topology of the compact space: for large $k$, 
the number of eigenvalues $N(k) \equiv\#\{j|k_j<k\}$ up to a given value
$k$, in an $n$ dimensional compact space
of volume $V_{\!\cal M}$, is given by~\cite{chav84}
\begin{equation} 
N(k) \equiv\#\{j |k_j<k\} \approx c(n)V_{\!\cal M} k^{n} + {\cal
O}(k^{(n-1)})\, ,~~~~~~~~~c(n)= V(unit~ball)/(2\pi)^n \, .
\lbl{weyl}
\end{equation} 
The constant $c(n)$ is related to the volume
within a unit sphere, set by the geometry of the space. In eq.(\ref{weyl}), 
$j$ counts the eigenvalues, including multiplicity. As a
corollary, the eigenvalues asymptotically can be
estimated:
\begin{equation} 
k_j \sim c(n)^{-1/n} (j/V_{\!\cal M})^{1/n}\ {\rm as} \ j\to \infty \, .
\lbl{weyl_cor}
\end{equation}
The corrections to Weyl's asymptotic formula are ${\cal
O}(k^{(n-2) + 2/(n+1)})$ for flat tori and ${\cal
O}(k^{(n-1)}/\ln(k^2))$ for manifolds with negative curvature.

Weyl's formula also points to the dependence of the infra-red cut off
on size: the smaller the space, the larger  the infra-red
cutoff. Eq.~(\ref{weyl_cor}) shows that in the limit of large
eigenvalues the typical spacing between distinct eigenvalues is given
by the inverse of its linear size. These are familiar and intuitive
facts in Euclidean compact spaces such as the simple tori.  Weyl's
formula encourages the view that these broad features in the spectra
of the compact manifolds transcend the geometry.  Another important
point of Weyl's formula relevant for our work is that the cumulative
number of eigenstates asymptotically depends only on the volume of the
manifold and not on its exact topology.  Thus, the gross properties of
the spectrum are shared by spaces with comparable volumes, and
quantities which are fairly democratically weighted integrals of the
power spectrum can be expected to be similar.

\section{Correlations in Compact Hyperbolic spaces}
\label{sec_res}
 
As a prelude to our computation of CMB temperature anisotropy
correlations for CH spaces in~\cite{paperB}, we focus our attention in
this section on the $\Phi$-correlation function $\xic$ between points
${\bf x}$ and ${\bf x}^\prime$ that belong to a $2$-sphere around the
origin in ${\cal M}^u$ --- \ie there exist discrete motions $\gamma,
\gamma^\prime \in \Gamma$ such that the points $\gamma {\bf x},
\gamma^\prime {\bf x}^\prime \in {\cal M}^u$ are equidistant from the
origin.

Correlation functions of this kind arise in evaluating large angle CMB
anisotropies associated with the gravitational potential on the sphere
of last scattering (SLS) -- the Naive Sachs-Wolfe (NSW) effect. (The
radius of the SLS, $R_{\sc LS} \approx 2 \tanh^{-1}\sqrt{1
-\Omega_0}$, is related to the density parameter, $\Omega_0$.)  The
angular correlation $C(\hat q,\hat q^\prime)$ between the NSW CMB
anisotropy in two directions $\hat q$ and $\hat q^\prime$ is given by
\begin{eqnarray}
 C(\hat q,\hat q^\prime)\equiv
\left\langle\dT(\hat q)\dT(\hat q^\prime)\right\rangle
&&{}= \frac{1}{9} 
\langle\Phi(\hat q\chiH,\tauls)\Phi(\hat q^\prime\chiH,\tauls)\rangle~.
\lbl{cthetaNSW}
\end{eqnarray}
In simply connected universes, $C(\hat q,\hat q^\prime) \equiv C(\hat
q\cdot\hat q^\prime)$ is statistically isotropic. In contrast, for all
compact universe models with Euclidean or Hyperbolic geometry, $C(\hat
q,\hat q^\prime)$ is statistically anisotropic. The breakdown of
isotropy leads to characteristic patterns in the predicted CMB
anisotropy determined by the shape of the Dirichlet domain around the
observer. Moreover, except for the simple flat torus, the global
inhomogeneity implies that the variance $C(\hat q,\hat q)$ varies with
direction $\hat q$. This implies that the CMB sky would be a
realization of an inhomogeneous field that would have characteristic
patterns of `loud' and `quiet' regions. These two effects
constitute two aspects of the signature of a compact universe.

There are three distinct regimes for the correlation patterns on the
sky in a compact space. For $R_{\sc LS} > R_>$, the patterns are
mainly dominated by the mapping of the SLS into the compact space. The
NSW-CMB patterns are characterized by spikes of positive correlation
when the neighborhood of one point on the SLS is multiply-imaged on
the SLS. For $R_{\sc LS} \ll R_<$, \ie the SLS is well within a single
domain, the compact space is indistinguishable from a simply-connected
space with the same geometry. In the intermediate $R_< \lta
R_{\sc LS} \lta R_>$ regime, there is very little multiple imaging; 
nevertheless, the SLS is large enough to feel the compactness of the
space. Typically the correlation pattern retains the structure of
correlation in the simply connected space, but is significantly
deformed.

\begin{figure}[h]
\plotone{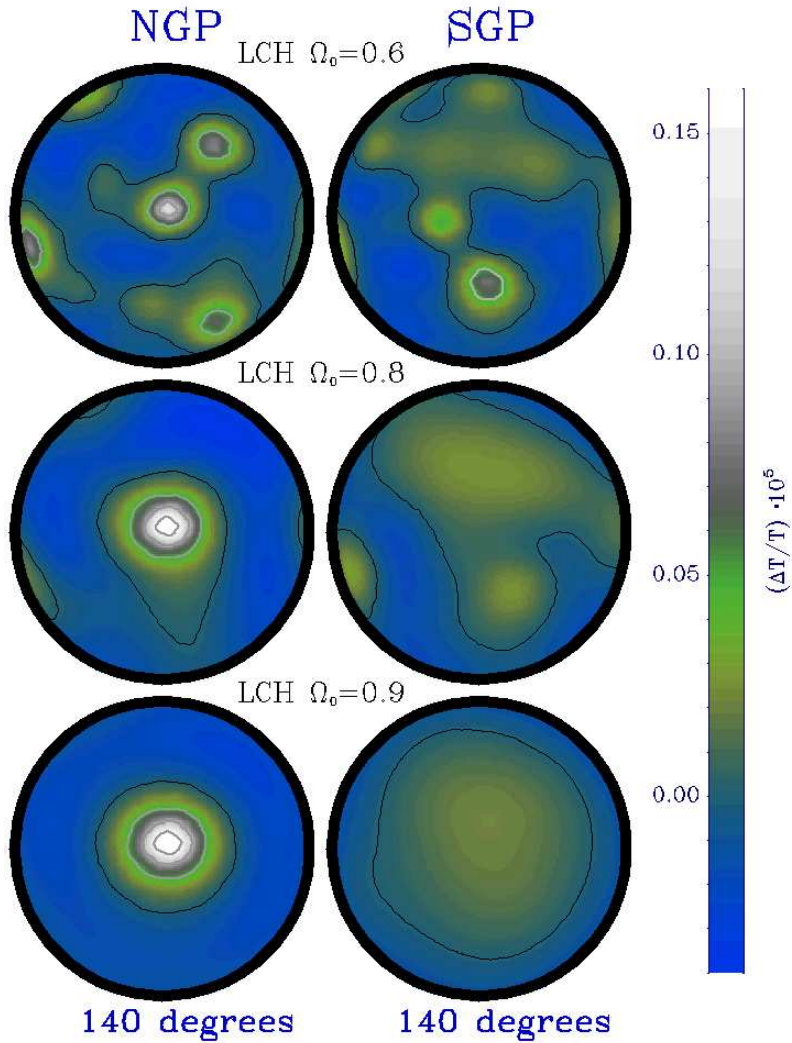}
\caption{Typical angular correlation patterns $C(\hat{q}_0 ,\hat{q})$
of the naive (or surface) Sachs-Wolfe effect on CMB anisotropy around
a fixed direction $ \hat{q}_0$ in the L(arge)CH model are shown as
full-sky maps at the angular resolution of $5.2^\circ\times5.2^\circ$
pixels. The full-sky maps are plotted as pairs of $180^\circ$ diameter
hemispherical caps, one centered on the South Galactic Pole (SGP) and
one on the North (NGP). $ \hat{q}_0$ points to the NGP.  In a simply
connected universe, the contours of equal correlation would be
concentric circles around $ \hat{q}_0$, due to the global
isotropy. The three values of $\Omega_0$ are representative of the
three regimes. In the top panel, the radius of the SLS, $R_{\sc LS}$,
is greater than $R_>$ and one sees multiple imaging. In the second,
with $R_< < R_{\sc LS} < R_>$, there are significant distortions but
no multiple imaging. The bottom panel has $R_< \ll R_{\sc LS}$, and no
observable correlation signatures of  compactness. In all the maps,
the dipole component of the correlation function has been
subtracted. The maps have also been smoothed by a $1.66^\circ$
Gaussian filter. }
\label{fig_corrcirc}
\end{figure}

Figure~\ref{fig_corrcirc} illustrates the typical correlation pattern
$C({\hat q_0}, {\hat q})$ in the large CH model v3543(5,1) around a
fixed direction, ${\hat q_0}$, which we chose to correspond to a
fiducial \bqt North Galactic Pole\eqt (NGP). Of course, the global
anisotropy implies that this pattern will differ for different choices
of ${\hat q_0}$. At $\Omega_0=0.6$, the SLS is larger ($R_{\sc
ls}=1.44$) than the domain ($R_>=1.32$), and spikes of enhanced
correlation are seen with widely separated directions when the fixed
point on the sphere, or points physically close to it, are multiply
imaged on the SLS.  At $\Omega_0=0.8$, the SLS is smaller ($R_< <
R_{\sc ls}=0.93 < R_>$) and high correlation spikes due to multiple
imaging are absent. Nevertheless, the compactness of the space is
evident in the distorted contours around ${\hat q_0}$. At
$\Omega_0=0.9$, the SLS is completely contained within the domain, and
the correlation around the NGP is circularly symmetric. As expected,
in this regime the compactness of the space on scales much larger than
the horizon has very little observational signal. The contours show
only slight, observationally insignificant, distortions.

\begin{figure}[h]
\plotone{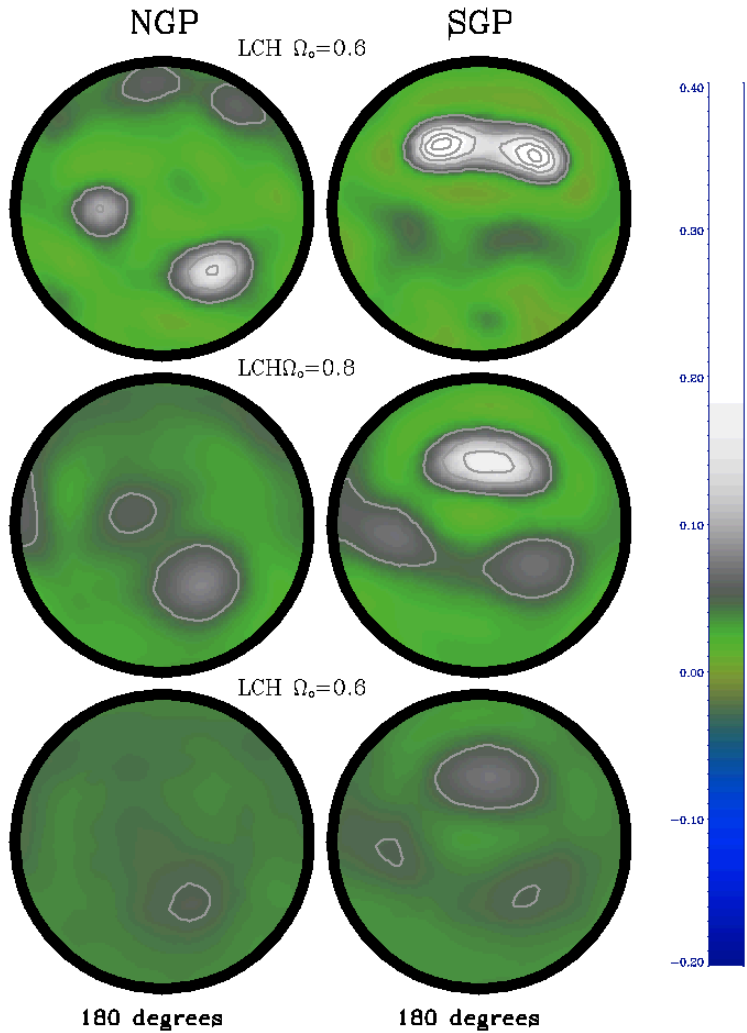}
\caption{The variations in the variance $C(\hat{q} ,\hat{q})$ of a
scalar field on three concentric spheres around the same
basepoint. The radius of the spheres correspond to $R_{\sc LS}$ at the
values of the density parameter $\Omega_0$ shown.  The variance is
shown in full-sky maps plotted as pairs of $180^\circ$ diameter
hemispherical caps, one centered on the South Galactic Pole (SGP) and
one on the North (NGP).  In addition to the \cobedmr beam smoothing,
the maps have been smoothed by a $1.66^\circ$ Gaussian filter. The
loud feature seen at $\Omega_0=0.8$ ($R_< < R_{\sc LS} < R_>$)
corresponds to $\approx 100\%$ excess in the variance over its mean.
At $\Omega_0=0.6$, the spot is multiply-imaged since the sphere is
larger than the domain $R_{\sc LS} > R_>$. At $\Omega_0=0.9$ there is
very little variation over the small sphere $R_{\sc LS} \ll R_<$. }
\label{fig_varvar}
\end{figure}

Figure~\ref{fig_varvar} plots the variance in the NSW-CMB temperature,
$C(\hat q,\hat q)$, in the large CH model v3543(5,1) for three values
of $\Omega_0$. The first two maps show a significant loud feature at
$\Omega_0=0.6$ and $\Omega_0=0.8$, corresponding to the radii $R_{\sc
LS}=1.44$ and $R_{\sc LS}=0.93$, respectively. The loud spot on the
sphere at $\Omega_0=0.8$ is within the domain. At $\Omega_0=0.6$, the
sphere is larger than the domain ($R_>=1.32$), and the loud spot is
multiply imaged on the sky.  The third map has $\Omega_0=0.9$,
corresponding to a sphere of radius $R_{\sc LS}=0.63$, which is
significantly smaller than the inradius, $R_< = 0.89$. The
variance does not show much variation over this small sphere around
the observer.

The CMB anisotropy has contributions other than the NSW term in
eq.~(\ref{cthetaNSW}). In particular, there is an integrated
Sachs-Wolfe contribution (ISW), with the integration along the line of
sight from the SLS to the observer.  The NSW term dominates when the
density parameter $\Omega_0$ is close to unity. When the ISW
contribution is important, it significantly modifies the correlation
patterns, as discussed in our companion paper~\cite{paperB}.

\section{Discussion}
\label{sec_dis}

The computation of the angular correlation function of the CMB
anisotropy requires machinery to compute spatial correlation functions
on equal-time spatial 3-hypersurfaces. We have presented a general
method of calculating these for multiply connected spatial sections
which evades the task of eigenmode decomposition. This is particularly
useful when considering compact hyperbolic models of the universe for
which rather little is known about the spectrum of the Laplacian, and
eigenmode decomposition is known to be difficult to obtain.

We summarize here the basic knowledge we require of a given manifold
to use our method and the steps to be followed to obtain a given
accuracy level for the correlation functions.  

 For the selected
compact model ${\cal M}={\cal M}^u/\Gamma$ of interest, we must be
able to construct the tiling of the universal cover ${\cal M}^u$ from
the generators of group $\Gamma$ (\S~\ref{sec_top}) up to some distance $
r_* $ from the origin, which will be determined by the required level
of accuracy.  This is in itself a daunting mathematical task, but,
fortunately, for many thousand compact hyperbolic models, the Snappea
package~\cite{Minn} gives enough information for us to carry out this
step.  

 Given the tiling, we perform our correlation function calculation
in a sequence of radial shells of size $r$, testing whether a stopping
criterion based on a desired level of convergence in the regularized
summation over images (\S~\ref{ssec_CH_numer}) is satisfied; if so, this
defines $r_*$. It may be that the required $r_*$ is beyond the
computational power at hand. For the correlation function calculations
of interest for the CMB problem in the compact hyperbolic spaces we
have tried, $10^5$ images are computationally very feasible (less than
a day on a 433 MHz alpha workstation). This allows us to go out to
$r_* \approx 5 d_{\!\cal M}$, more than adequate for convergence.  

 It may be for manifolds with very many faces, the number of images
required to converge could be prohibitive. Even if this is so, we
obtain useful results out to the radius we can achieve because at each
shell $r$ the symmetries of the manifold are preserved in the
correlation function. Indeed even the nearest few shells of images are
enough for a qualitatively correct result on the basic pattern of
correlations, which is also quantitatively not too bad.  We showed
explicitly for the flat torus (\S~\ref{sec_tor}) that the correlation
function is well determined as long as the density of states
calculated for that number of shells has the power in each discrete
mode localized to within $\Delta k_i \sim 1/d_{\!\cal M}$ around the
true eigenvalue $k_i$. Since the line of argument leading to this
result is independent of the local geometry of the compact space, it
applies to CH spaces as well. 

 Our method has the merit of avoiding explicit eigenmode
decomposition. Recently progress has been reported in directly
computing the low-$k$ eigenmodes of the Laplacian for selected compact
hyperbolic spaces~\cite{eigen1}. It has proved necessary to use
spectral line ``deblending'' techniques in conjunction with these
methods to get the eigenvalue spectrum accurately~\cite{eigen2}. As we
showed in \S~\ref{ssec_powspec}, we can also use the method of images
to calculate the spectrum of eigenvalues, though this is more
difficult than correlation function evaluation since it requires
longer summations to get narrow spectral lines, as is evident from
Fig.~\ref{fig_tor_moi2} and \cite{us_cwru}. These figures immediately
suggesting deblending, but at this stage, only for the torus case do
we explicitly know the shapes of the broadened spectral lines obtained
by sequential image addition. One of the manifolds in which the
low-$k$ eigenvalues were determined by Inoue~\cite{eigen1} is in common with
those we have computed the spectrum for. We find his results agree
with ours.
 
The goals of this paper were to provide a detailed description of our
regularized method of images, demonstrate that it works extremely
well, and more generally show the basic effects of compactness on the
correlation function of a scalar field and the density of states of
the Laplacian. The computation of large angle CMB anisotropy, the
features that arise in the CMB correlation that are characteristic of
compact spaces, the detailed comparison with the \cobedmr data and the
constraints that follow using large and small CH space examples are
presented in the companion paper~\cite{paperB}. A compilation of the
constraints from the all sky \cobedmr data on a large selection of CH
spaces will be presented in~\cite{us_prl}.

\section{Acknowledgements}
We acknowledge the use of the SnapPea package and related material on
compact hyperbolic spaces available at the public website of the
Geometry center at the University of Minnesota. TS acknowledges
support during the final stages of this work from NSF grant
EPS-9550487 with matching support from the state of Kansas. In the
course of this project we had enjoyable interactions with J.~Levin,
N.~Cornish, D.~Spergel, I.~Sokolov, G.~Starkman and J.~Weeks.

\appendix

\section{Inflationary perturbations in a Hyperbolic universe}
\label{app_genpert}

It is widely accepted that the large scale structures in the present
Universe grew out of small initial metric perturbations via the
mechanism of gravitational instability. The inflationary epoch in the
Universe provides us with a setting for the generation of fluctuations
with power on large scales. Cosmological perturbations are effectively
massless (light) scalar fields residing on a background FRW space-time
and the large scale power is related to the infra-red behavior of the
light scalar field propagator in the inflationary epoch.\footnote{This
infra-red behavior is linked to the infra-red divergence which
strictly exists for massless scalars and for infinite duration of the
accelerated phase. In the case of inflationary scenarios, both the
finite duration and small effective mass serve to regulate the
divergence.} In this Appendix, we present a simplified calculation of
the initial power spectra for open inflation (hyperbolic geometry) to
place the spectrum that we use in our work within a broad class of
``tilted'' inflationary spectra which are ``scale free'' on scales
much smaller than $d_c$ and are analogous to the well-known tilted
spectra in Euclidean (flat) inflationary universe. The scope of our
analysis is limited by the particular choice of the vacuum state,
as discussed below.

Inflationary scenarios that lead to a simply connected hyperbolic
universe generally require two stages~\cite{openbub}. This is because
a single stage of inflation cannot provide homogeneity across the
Hubble patch without inflating the local curvature in the patch to
negligible values. A widely prescribed solution is to invoke a
``creation stage'' involving the creation of a universe with
homogeneous hyperbolic spatial sections which is then followed by a
sufficiently short inflationary stage (so as not to expand the
curvature away) inflationary stage responsible for generating the
primordial perturbations at cosmological scales. Since the second
epoch of inflation cannot be long, the specifics of the first stage of
creation of $\hm$ spatial sections could influence the spectrum of
generation of perturbations at supercurvature scales. The possible
effects on the spectrum have been extensively studied in recent
literature~\cite{openbubspec} in the context of open-bubble models in
which the open universe resides in the interior of the bubble
nucleated in the first-order phase transition from a larger inflating
universe. It is doubtful, however, that the bubble mechanism can be
used to produce compact hyperbolic spaces.  More relevant is probably
the mechanism of quantum creation of the universe ``from nothing''
where the topology of the created space is related to the topological
properties of the instanton solution in the Euclidian
section~\cite{CHcreat}.  Proper quantization of the fluctuations in
such a scenario is, however, still a matter for the future.

We assume that at the beginning of the second (inflationary) stage all
quantum fields on the hyperbolic hypersurface are in the vacuum state
and the vacuum modes in eq.~(\ref{posvmode}) are chosen by identifying
the positive frequency part of the general solution to the mode
evolution equation.  This choice of the initial state sets the
boundary of applicability for our considerations
\footnote{
Although not explicitly used, the presence of the ``creation stage'' is
essential. On the hyperbolic chart of the DeSitter space some
observable $1$-loop quantities computed in this vacuum would diverge
on the ($t=0$) boundary and hence the conditions assumed here at the
onset of the inflationary epoch must breakdown in the past,
presumably, due to the creation stage.}. It is important
to bear in mind that, in general, the creation stage could modify the 
proper choice of the vacuum state and consequently the predicted spectrum
eqs.~(\ref{spec_opentilt}) and (\ref{spec_openHZ}).

We outline the derivation of the power spectrum for a free scalar
field $\psi$ which will be appropriately identified with metric
perturbations subsequently.  The scalar field equation allows the
spatial and temporal dependence to be separated: 
$\psi({\bf x}, \tau) = \psi_k(\tau) ~G_{\bf k}({\bf x})$.  The rapid
expansion of the space-time during inflation would redshift away to
insignificance the number density of any pre-inflationary quanta
within the first few e-folds of expansion. As a result, it is a good
approximation to assume that all the fields are in their vacuum state
i.e., $\psi_k(\tau) \equiv \psi^{(+)}_k(\tau)$.  The vacuum
expectation value of the field, $\langle\psi^2(\tau)\rangle$, is the
coincidence limit (${\bf x}' \to {\bf x}$) of the equal-time two-point
function, $\langle\psi({\bf x}', \tau) \psi({\bf x}, \tau)\rangle$,
and, hence, can be expressed as a mode-sum over the mode functions:
\begin{eqnarray}
\langle\psi^2\rangle &=& \frac{1}{(2\pi)^3}~\int
d^3k~\psi_k(\tau)~\psi^{*}_k(\tau) \nn \\
&\equiv& \int \frac{dk}{k}~
\left[{k^3\over2\pi^2}~\psi_k(\tau)~\psi^{*}_k(\tau)\right]~.
\lbl{specdef1}
\end{eqnarray}
The quantity within the square brackets represents the contribution
per logarithmic $k-$ interval to the total power at a given time,
$\tau$. We usually define the power spectrum by evaluating this
quantity at some convenient time, $\tau_{*}$, during inflation. For
instance, $\tau_{*}$ could correspond to the time during the
inflationary epoch when the comoving scale, $k_0$, corresponding to
the observed horizon was equal to the Hubble radius ($k_0 |\tau_{1}| =
k_0\tau_0$). The power spectrum, $P_\psi(k/k_0)$, of the field,
$\psi$, is defined to be
\begin{eqnarray}
P_\psi(k/k_0) &\equiv &  {k^3\over2\pi^2}~\psi_k(\tau_{1})
~\psi^{*}_k(\tau_{1})~.
\lbl{specdef2a}
\end{eqnarray}
The present astrophysical comoving scales correspond to physical
scales which were much larger than the Hubble radius at the end of
inflation, hence it suffices to evaluate the amplitude of modes in the
expression for the power spectrum, $P_\psi(k/k_0)$, in the limit,
$k\tau \to 0$.

We now compute the spectrum of initial perturbations generated in the
second inflationary stage on a FRW model with $\hm$ spatial sections.

The spatial modes of a scalar field on $\hm$ are given by orthonormal
eigenfunctions of the Laplacian in spherical coordinates:\cite{abb_sch86}
\begin{eqnarray} 
&&G_{\bf k}({\bf x}) = \sum_{\ell=0}^\infty \sum_{m=-\ell}^{m=\ell}
{\cal G}_k^\ell(r)\,\,Y_{lm}(\hat x);~~~~~~~~ {\cal G}_k^\ell(r) = \sqrt{\frac{N_\ell(\beta)}{\sinh
r}}\,\,\pleg{-\ell-\half}{-\half+i\beta}{\cosh r} ,\nn\\ &&\beta =
\sqrt{|{\bf k}|^2 -1},\,  r=|{\bf x}|, ~~~\hat x = {\bf x}/r,~~~
N_\ell(\beta)= \bigg|\frac{\Gamma(i\beta + \ell+1)}{\Gamma(i\beta+1)}\bigg|^2.
\lbl{open_spmodes}
\end{eqnarray}
In the above equation, the $\pleg{\mu}{\nu}{x}$ are the associated
Legendre functions and $\Gamma(\nu)$ is the Euler-Gamma
function~\cite{AbStg}.  Owing to the isotropy of $\hm$, the modes
depend only $k=|{\bf k}|$. Moreover, modes with $k d_c\ge 1$, (\ie,
$\beta$ real and positive) form a complete orthonormal basis for free
fields on $\hm$, and it is convenient to use the wavenumber, $\beta$,
to label these modes.

The temporal modes of the scalar field, $\ipertb$, on a hyperbolic
universe, obey the equation

\begin{equation}
\ipertb\eqt + 2 \frac{a'}{a} \ipertb'
+ \left[ (\beta^2+1)+ a^2 m_{\rm eff}^2\right]\ipertb =0 \, , 
\lbl{eq_ipert}
\end{equation}
where $m_{\rm eff}^2$ is the effective mass of the field and the prime denotes
density derivatives with conformal time.  We now
consider the scalar field $\psi$ to be the fluctuations, $\d\phi$, of
the inflaton field~\cite{mukh_92,bonLH} and then $m_{\rm eff}^2$ is
related to the quantities set by the evolution of the background
inflaton field and the metric.

The above setting allows us to investigate quantum fluctuations during
inflation in some well defined regimes. The classification here
closely follows that of inflation in a flat universe.  We consider the
following cases: 

\begin{itemize}

\item[i.] uniform inflationary (deSitter) expansion and $m_{\rm
eff}^2=0$. This is the analog of standard slow-roll inflation;

\item[ii.] uniform inflationary expansion with a constant value of
$m_{\rm eff}^2/H^2\neq 0$. This is the analog of the class of models
with inverted harmonic oscillator like potentials, such as natural
inflation; and

\item[iii.] constant equation of state during inflation. This is an
abstraction of the power-law models of inflation.

\end {itemize}
In each of the above, the conditions are meant to hold in an
approximate fashion over the relevant range of astrophysical scales,
just as for the flat inflation analogs.

In the case of deSitter expansion, the scale factor is $a= \sinh(Ht)/H=
-\cosech\tau/H$, expressed as a function of cosmic time $t$ and conformal
time $\tau$. Eq.~(\ref{eq_ipert}) then reduces to
\begin{equation}
\ipertb\eqt - 2 \coth\tau \, \ipertb'
+ \left[ (\beta^2+1)+ (m_{\rm eff}/H)^2/\sinh\tau \right]\ipertb =0 \, .
\lbl{eq_ipert1}
\end{equation}
The general solution involves associated
Legendre functions $\qleg{-\mu}{-\half+i\beta}{\cosh\tau}$ and
$\pleg{-\mu}{-\half+i\beta}{\cosh\tau}$, multiplied by a factor
$\sinh^\frac{3}{2}\tau$. The index $\mu=\sqrt{9/4-m^2_{\rm eff}/H^2}$.
The positive frequency (vacuum mode) solution is identified with the
late time asymptotics $\sim e^{-ik\tau}$. The normalized vacuum mode
solution, $\ipertb^{(+)}$,  is 
\begin{equation}
\ipertb^{(+)} = H^2\sqrt{\frac{\Gamma(\half+\mu+i\beta)}
{i\Gamma(\half-\mu+i\beta)}}
\sinh^2\tau \,\,\qleg{-\mu}{-\half+i\beta}{\cosh\tau} \, , 
\lbl{posvmode}
\end{equation}
where the overall normalization is determined by normalizing the
probability current, \ie setting the Wronskian of $\ipertb^{(+)}$
equal to $i a^{-2}$.

The spectrum of initial perturbations in the inflaton field can be
calculated from the vacuum mode solution through
eq.~(\ref{specdef2a}):
\begin{equation}
P_{\ipert}(\beta)= \left|i \frac{\Gamma(\half-\mu+i\beta)}
{\Gamma(\half+\mu+i\beta)}\right|^2 \,.
\lbl{spec_opentilt}
\end{equation}
If $m^2_{\rm eff}=0$, the spectrum of initial perturbations in the
inflaton field reduces to

\begin{equation}
P_{\ipert}(\beta)= \frac{1}{\beta(\beta^2+1)}.
\lbl{spec_openHZ}
\end{equation}
This is the analog of the Harrison-Zeldovich spectrum for hyperbolic
models derived in \cite{lyt_stew90,rat_peeb94}. In the limit $\beta
\gg 1$, the spectrum goes over to the well-known flat \bqt flicker noise\eqt
result. This is also the power spectrum which implies equal power per
logarithmic interval in $|{\bf k}|$ that we use in our work on CH
spaces (see $\S$\ref{ssec_hypcorr}). 

The power spectrum in eq.~(\ref{spec_opentilt}) is best numerically
computed by using Lancoz's formula for Gamma functions. Using the
Stirling approximation to Gamma functions for $\beta \gg 1$, one 
recovers the standard tilted flat space spectrum from
eq.~(\ref{spec_opentilt}), ${\cal P}_\Phi \sim k^{3/2-\mu}$.

\begin{figure}
\plotone{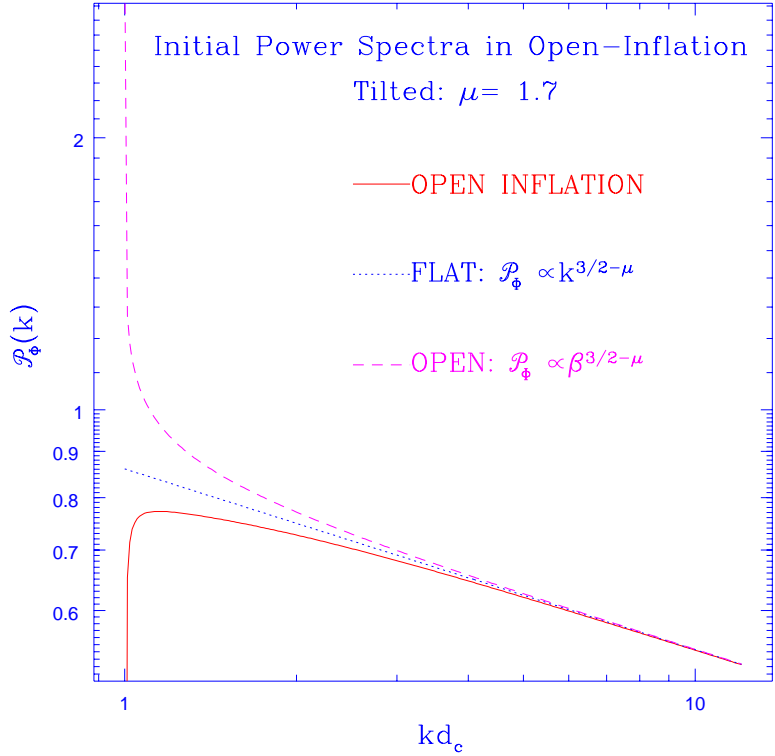}
\caption{ The initial power spectrum expected from inflation in a
hyperbolic universe with a small tilt is shown as the solid curve
(OPEN INFLATION). There is a sharp cutoff below $kd_c=1$ since the
quantum fluctuations during inflation cannot excite supercurvature
modes. The curve labeled FLAT shows the analogous tilted spectrum in a
flat universe. The open inflation and the flat curves match at
$kd_c\gg 1$. Tilted initial spectra in Hyperbolic universes have often
been taken to be a power law in $kd_c$ or $\beta$. Although equivalent
at $kd_c\gg 1$, these differ at small $kd_c$ from the inflationary
prediction; the former is labeled FLAT while the latter is depicted by
the dashed curve labeled OPEN. The open inflation spectrum that we
use in our paper corresponds to no-tilt, $\mu=3/2$, is flat with a
sharp cutoff at $kd_c=1$.}
\label{fig_openspec}
\end{figure}

In the more general case of constant equation of state, $w\equiv
P/\rho$, during inflation, the scale factor obeys $a =
\sinh^\alpha(\tau/\alpha)$ where $\alpha=2/(3w+1)$.  The deSitter
expansion corresponds to the case $w=-1$.  The eq.(\ref{eq_ipert}) for
the temporal modes reduces to the form

\begin{equation}
\ipertb\eqt + 2 \coth(\tau/\alpha)\, \ipertb'
+  (\beta^2+1)\ipertb =0 ~.
\lbl{eq_ipert2}
\end{equation}
The general solution of the above equation involves the same
associated Legendre functions as the solution for equation
(\ref{eq_ipert1}) with index $\mu=1/2-\alpha$, rescaled wavenumber
$\tilde\beta^2 =\alpha^2\beta^2 = (k^2-1)\alpha^2$, rescaled time
$\tilde\tau =\tau/\alpha$, and multiplication by a factor of
$\sinh^\mu\tilde\tau$.

The positive frequency (vacuum mode) solution is identified again with
the late time asymptotics $\sim e^{-ik\tilde
\tau}$. The normalized vacuum
mode solution, $\ipertb^{(+)}$,  is 
\begin{equation} 
\ipertb^{(+)} =
H^2\sqrt{\frac{\Gamma(\half+\mu+i\tilde\beta)}
{i\Gamma(\half-\mu+i\tilde\beta)}}
\sinh^2\tau \,\,\qleg{-\mu}{-\half+i\tilde\beta}{\cosh\tilde\tau}.
\lbl{posvmode2}
\end{equation}
The spectrum of perturbations is of the same form as
(\ref{spec_opentilt}) with rescaled wavenumber $\tilde\beta$ and
$\mu=1/2-2/(3w+1)$. The spectra given by (\ref{spec_opentilt}) are
hyperbolic universe analogs of tilted spectra generated in flat
universe inflation models such as power-law inflation.

\end{document}